\documentclass[twocolumn]{aastex701}

\begin{document}

\title{Constraining the Merger Fraction and the Upper Mass Limit of Ultramassive White Dwarfs}

\author[0009-0009-9105-7865]{Gracyn Jewett}
\affiliation{Homer L. Dodge Department of Physics and Astronomy, University of Oklahoma, 440 W. Brooks St., Norman OK, 73019, USA}
\email[show]{gjewett@ou.edu}

\author[0000-0001-6098-2235]{Mukremin Kilic} 
\affiliation{Homer L. Dodge Department of Physics and Astronomy, University of Oklahoma, 440 W. Brooks St., Norman OK, 73019, USA}
\email{kilic@ou.edu}

\author[0000-0003-2368-345X]{Pierre Bergeron} 
\affiliation{Département de Physique, Université de Montréal, C.P. 6128, Succ. Centre-Ville, Montréal, QC H3C 3J7, Canada}
\email{pierre.bergeron.1@umontreal.ca}

\author[0009-0003-3096-7618]{Alex Gleason} 
\affiliation{Homer L. Dodge Department of Physics and Astronomy, University of Oklahoma, 440 W. Brooks St., Norman OK, 73019, USA}
\email{Alexander.R.Gleason-1@ou.edu}

\author[0000-0002-4462-2341]{Warren R.\ Brown}
\affiliation{Center for Astrophysics, Harvard \& Smithsonian, 60 Garden Street, Cambridge, MA 02138, USA}
\email{wbrown@cfa.harvard.edu}

\author[0000-0002-6153-9304]{Manuel Barrientos}
\affiliation{Homer L. Dodge Department of Physics and Astronomy, University of Oklahoma, 440 W. Brooks St., Norman OK, 73019, USA}
\email{mbarrientos@ou.edu}

\author[0000-0002-0948-4801]{Matthew J. Green}
\affiliation{Homer L. Dodge Department of Physics and Astronomy, University of Oklahoma, 440 W. Brooks St., Norman OK, 73019, USA}
\email{matthew.j.green-2@ou.edu}

\author[0000-0002-9632-1436]{Simon Blouin}
\affiliation{Department of Physics and Astronomy, University of Victoria, Victoria BC V8W 2Y2, Canada}
\email{sblouin@uvic.ca}

\author[0000-0003-4236-9642]{Vikram S. Dhillon}
\affiliation{Astrophysics Research Cluster, School of Mathematical and Physical Sciences, University of Sheffield, Sheffield, S3 7RH, UK}
\affiliation{Instituto de Astrof\'{\i}sica de Canarias, E-38205 La Laguna, Tenerife, Spain}
\email{Vik.Dhillon@sheffield.ac.uk}

\author[0000-0001-7221-855X]{Stuart Littlefair}
\affiliation{Astrophysics Research Cluster, School of Mathematical and Physical Sciences, University of Sheffield, Sheffield, S3 7RH, UK}
\email{s.littlefair@sheffield.ac.uk}

\begin{abstract}
We performed a spectroscopic survey of all known white dwarfs with $M \geq 1.3~M_\odot$ and $T_{\rm eff}\geq11,000$ K identified through Gaia DR3 astrometry from the Gentile-Fusillo et al. (2021) catalog. Of the 293 white dwarf candidates that met our criteria, 259 had no optical spectroscopy or spectral classification available in the literature. We obtained follow-up spectroscopy on 97\% of this sample and performed a detailed model atmosphere analysis to constrain the atmospheric composition and mass for each object. For a selection of magnetic white dwarfs, we use offset dipole models to constrain the magnetic field strength and geometry and calculate the magnetic fraction as a function of mass. We use magnetism, unusual atmospheric composition, kinematics, and rapid rotation rates to identify merger remnants, and constrain the merger fraction to be $53_{-6}^{+7}$\% for white dwarfs more massive than $1.3~M_\odot$. Hence, roughly half of the most massive white dwarfs likely form through binary mergers. We also use our sample to derive an empirical upper mass limit for white dwarfs in the solar neighborhood. Including the General Relativistic effects, the most massive white dwarf in our sample, J0318+7923, has $M=1.33-1.35~M_\odot$ depending on the unknown core composition. Furthermore, because our temperature criteria overlap with the ZZ Ceti instability strip, we find a new selection of ultramassive pulsator candidates that may provide an excellent opportunity to probe the interiors of crystallized CO or ONe-core white dwarfs.  
\end{abstract}


\section{Introduction}
White dwarfs are the final evolutionary state for the majority of stars \citep{fontaine01}. Single
white dwarfs can form in isolation, but they can also form through stellar mergers during the main-sequence, post-main-sequence, or the white dwarf phase \citep[e.g.,][]{toonen17}.
Binary population synthesis models predict that 30 to 50\% of massive ($M \geq 0.9~M_\odot$), single white dwarfs form through a merger event. The evolutionary channel that is dominant for massive white dwarfs are mergers of double white dwarf systems contributing $\sim45$\% of the events \citep{temmink20}. 

Determining whether a massive, single white dwarf is the product of a binary merger is not trivial, but there are observational signatures that may suggest a merger origin. These signatures are unusual atmospheric composition, a strong magnetic field, rapid rotation, and unusual kinematics.    
The first category of merger remnants with unusual atmospheric composition includes warm/hot DQ white dwarfs with carbon rich atmospheres \citep{dufour08,koester19,kilic25,ouldrouis26}, high mass ($M\gtrsim1~M_\odot$), and unusual kinematics, all of which indicate a merger origin \citep{dunlap15supernova,coutu19,kawka23}.  

While there are a few theories on the origin of magnetism in massive white dwarfs, \citet{garcia12} showed that a strong magnetic field can be produced by the differentially rotating convective outer layers of a double white dwarf merger remnant. Due to the angular momentum of the initial binary system, merger products of white dwarfs are predicted to rotate rapidly, with rotation periods of $\sim$10-20 minutes \citep{schwab21}, much faster than the average mass white dwarfs formed via single star evolution \citep{hermes17}. Merger products can also have unusually high tangential velocities for their cooling age, which indicates there was a reset in the white dwarf's age due to a merger event, making the white dwarf appear younger than other white dwarfs with similar tangential velocities \citep{wegg12, cheng20}. 

Strong magnetic fields and rapid rotation have been shown to be characteristic of double white dwarf mergers \citep{garcia12, schwab21}. However, unusual kinematics and unusual atmospheric composition 
can also be produced through other formation channels, i.e. the formation of delayed Q branch white dwarfs
through mergers of subgiant + white dwarfs \citep{shen23}.
The strongest merger product candidates will show multiple signatures.

Empirical constraints on the white dwarf merger fractions of massive and ultramassive white dwarfs are scarce. Previous studies suffered from small sample sizes of white dwarfs with $M \geq 1.3~M_\odot$. For example, \citet{kilic23merger} had a sample size of 25 with a merger fraction of $56_{-10}^{+9}$\%, and \citet{jewett2024} had a sample size of 13 objects with $M \geq 1.3~M_\odot$ and a merger fraction of $38_{-11}^{+15}$\%. Constraining the merger fraction is key to understanding common envelope evolution and the contribution of double white dwarf mergers to the type Ia supernovae rate \citep{iben84}. For example, \citet{cheng20} found a double white dwarf merger fraction of $\sim20$\% among 1.2-1.3 $M_\odot$ white dwarfs. Based on these results and binary population synthesis models, they estimate a merger rate of $0.3\times10^{-13}~M_\odot^{-1}$ yr$^{-1}$ for super-Chandrasekhar double white dwarf mergers. They conclude that such mergers can explain a significant fraction of the supernovae Ia rate \citep[see also][]{maoz18,adamanePallathadka26}.

The white dwarf mass distribution in the solar neighborhood peaks at 0.59 $M_\odot$ with a tail towards higher masses \citep{kilic20sdss,obrien24}. 
The upper mass limit for white dwarfs is set by the electron degeneracy pressure, which can support stars up to a mass of $\sim1.4 M_\odot$ \citep{chandrasekhar31}. However, we do not expect single star evolution to produce white dwarfs more massive than about $1.37~M_\odot$,
as their 8-10~$M_\odot$ potential progenitors instead end up in electron capture supernovae \citep{nomoto87}. Binary evolution can lead to more
massive white dwarfs, and depending on the core composition (CO or ONe), those stars either explode as a Type Ia supernovae \citep{whelan73,iben84}
or collapse into neutron stars. Gaia DR3 provides an excellent opportunity to provide an empirical upper mass limit for white dwarfs in the solar neighborhood.

Here, we present the results from a spectroscopic survey of all white dwarfs with $M \geq 1.3~M_\odot$ from the \citet{gentilefusillo21} catalog with photometric $T_{\rm eff}\geq11,000$ K. We discuss sample selection and observations in Section \ref{selection and observations}. In Section \ref{model atmo}, we explain the model atmosphere analysis and present an in-depth discussion on each spectral type. Section \ref{mass dis} presents our hot and massive white dwarf sample used for the merger fraction calculation. Section \ref{search for merger signatures} presents the details of our search for merger signatures. Section \ref{discussion} discusses the merger fraction, the empirical upper mass limit for white dwarfs, and the ZZ Ceti instability strip. We state our conclusions in Section \ref{conclusions}.

\section{Sample Selection \& Observations}
\label{selection and observations}

Our two immediate goals are to constrain the merger fraction of ultramassive white dwarfs, and to find the most massive white dwarf in the solar neighborhood to provide an empirical upper mass limit for white dwarfs. To create a sample to answer these questions, we searched the \citet{gentilefusillo21} Gaia EDR3 catalog of white dwarfs for candidates with $M \geq 1.3~M_\odot$ and $T_{\rm eff}\geq11,000$ K based on pure hydrogen atmosphere solutions provided in that work. The mass criterion sets us with targets that are likely the most massive white dwarfs present, and the temperature criterion ensures that there is no ambiguity in the model atmosphere analysis as H, He, and C lines would be detected, if present. 

These criteria generated a sample of 293 ultramassive white dwarf candidates. At the time of our initial search, 259 had no spectral classification in the literature. There are 16 white dwarfs in common with \citet{jewett2024}, 23 with spectral types in the Montreal White Dwarf Database \citep[MWDD,][]{dufour17}, and two with spectra from the Dark Energy Spectroscopic Instrument \citep[DESI,][]{desicollab26}. We were not able to observe 7 objects in our sample due to telescope time and weather constraints, leaving our sample at 97\% complete. The observational parameters for all 293 targets from our initial search are shown in Table \ref{tab:short obs parameters}.

\begin{figure}
\hspace{-0.3in}
\includegraphics[width=3.6in, clip, trim={0in 0in 0.4in 0.4in}]{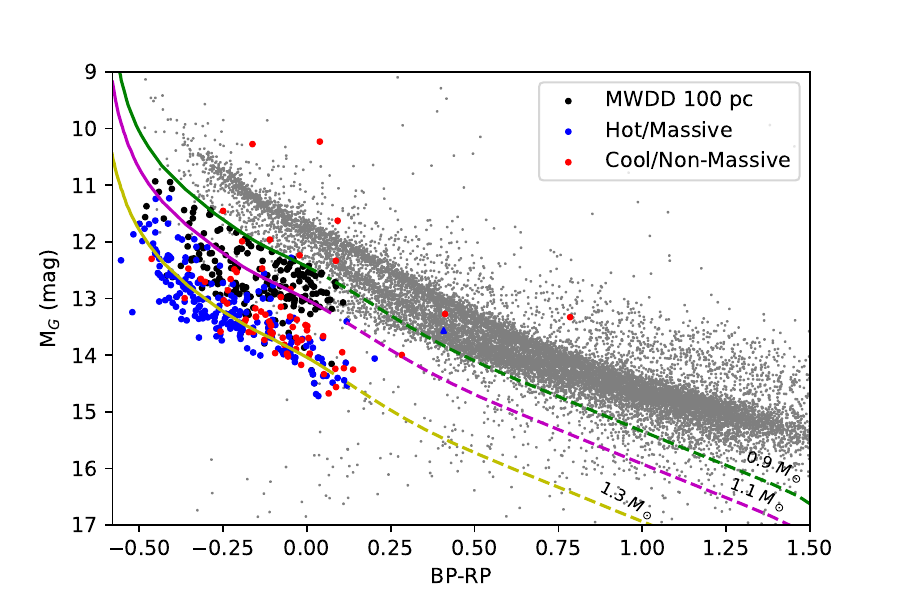}
\caption{Color magnitude diagram with evolutionary tracks for 0.9, 1.1, and $1.3~M_{\odot}$ pure H atmosphere white dwarfs. The solid and dotted tracks are for temperatures above and below 11,000 K, respectively. Candidates with photometric $M \geq 0.9~M_\odot$ and $T_{\rm eff}\geq11,000$ K (based on our analysis) are shown in blue, and the cooler or less massive targets are shown in red. Black points mark the massive MWDD 100 pc white dwarf sample of \citet{jewett2024}, and the gray points in the background show the 100 pc sample in the Pan-STARRS footprint. The blue triangle marks an outlier, J1509$-$4930, that has been removed from our massive and hot white dwarf sample.}
\label{fig:color mag}
\end{figure} 

\begin{deluxetable*}{crccccccccc}
\tablecolumns{10} \tablewidth{0pt}
\tabletypesize{\scriptsize}
\tablecaption{Observational properties of the initial white dwarf sample based on Gaia Data Release 3, with distances from \cite{bailerjones21}. This table is available in its entirety in machine-readable format in the online journal. A portion is shown here for guidance regarding its form and content. \label{tab:short obs parameters}}
\tablehead{\colhead{Object name} & \colhead{Gaia DR3 ID} & \colhead{RA} & \colhead{DEC} & \colhead{Distance} & \colhead{$\mu$$_{RA}$} & \colhead{$\mu$$_{DEC}$} & \colhead{G} & \colhead{$G_{BP}$} & \colhead{$G_{RP}$}\\
 & &  ($^{\circ}$) & ($^{\circ}$) & (pc) & (mas yr$^{-1}$) & (mas yr$^{-1}$) & (mag) & (mag) & (mag)}
\startdata
J0003+2603 & 2853237347003758080 & 0.94223 & 26.05677 & $277.5_{-41.3}^{+91.1}$ & 15.5 & $-$0.7 & 20.46 & 20.39 & 20.57\\ 
J0011$-$5953 & 4906000026775525504 & 2.82386 & $-$59.89067 & $144.1_{-2.5}^{+2.8}$ & 56.8 & 7.5 & 18.40 & 18.26 & 18.66\\ 
J0012+5708 & 422378456967056640 & 3.18384 & 57.14632 & $267.6_{-13.3}^{+12.6}$ & 14.7 & $-$10.7 & 18.76 & 18.65 & 19.10\\ 
J0018+2402 & 2849141528755983616 & 4.71408 & 24.04109 & $145.6_{-13.5}^{+15.5}$ & 71.4 & $-$43.9 & 20.38 & 20.49 & 20.40\\ 
J0028+0947 & 2751004824572504064 & 7.17004 & 9.78787 & $170.0_{-10.6}^{+11.6}$ & 33.1 & 5.4 & 19.59 & 19.50 & 19.75\\ 
J0043+8003 & 566346963315969408 & 10.97090 & 80.05519 & $160.1_{-4.5}^{+4.6}$ & $-$9.6 & $-$3.7 & 18.84 & 18.75 & 19.11\\ 
J0049$-$0810 & 2522225939783426560 & 12.27580 & $-$8.17000 & $123.6_{-3.4}^{+3.8}$ & 23.1 & $-$18.4 & 18.72 & 18.66 & 18.97\\ 
J0050$+$3138 & 360858960322547968 & 12.57920 & 31.64609 & $75.4_{-0.7}^{+0.8}$ & $-$90.1 & $-$38.7 & 18.08 & 18.06 & 18.17 \\
J0056$-$3347 & 5003385493930395776 & 14.10533 & $-$33.79510 & $97.3_{-0.8}^{+1.1}$ & $-$1.1 & $-$8.8 & 17.70 & 17.58 & 17.95\\ 
J0118$-$0156 & 2533306985471073920 & 19.54309 & $-$1.93677 & $100.2_{-0.8}^{+0.7}$ & 25.4 & $-$9.3 & 16.68 & 16.50 & 17.00 \\
\enddata
\end{deluxetable*}

Our observations took place at three telescopes. We used the Gemini South telescope with the Gemini Multi-Object Spectrograph (GMOS) to obtain optical spectroscopy of 111 targets. These observations were conducted as part of the Gemini queue programs GS-2023B-Q-329, GS-2024A-Q-331, GS-2024B-Q-319, GS-2024B-Q-405, GS-2025A-Q-240, and GS-2025B-Q-222. The 26 targets of the GS-2023B-Q-329 program were observed with the B600 grating and a 1" slit which provide a resolving power of $R = 844$. All other programs used the B480 grating and a 1" slit providing a resolving power of $R = 761$. The B480 grating was chosen over B600 due to its superior throughput and wavelength coverage.

We observed 110 targets at the 6.5 m MMT. We used the Blue Channel spectrograph with the 500 l mm$^{-1}$ grating and the 1.25" slit. This setup provides a 4.5 $\text{\AA}$ spectral resolution over $3850 < \lambda < 7000~\text{\AA}$ \citep{schmidt89}. For 4 targets, we used Binospec with the 270 l mm$^{-1}$ grating giving a 4.9 \text{\AA} spectral resolution. 
We observed 23 additional targets at the 6.5 m Magellan telescope with the MagE spectrograph. We used the 0.85" slit which provides a resolving power of $R = 4800$ and a wavelength range from 3400 to 9400 $\text{\AA}$.

Figure \ref{fig:color mag} shows the color magnitude diagram of our initial sample for the targets we have observed (red and blue points) as well as the massive MWDD 100 pc sample
of \citet[][black]{jewett2024} and the 100 pc white dwarf sample in the Pan-STARRS footprint (gray points in the background).  We include theoretical evolutionary tracks for 0.9, 1.1, and $1.3~M_{\odot}$ pure H atmosphere white dwarfs for comparison. Our selected candidates are on average fainter and therefore more massive than the \citet{jewett2024} sample. Even though our candidates were all selected
to have $M\geq1.3~M_\odot$ based on \citet{gentilefusillo21} model fits assuming a pure H atmosphere and using Gaia astrometry and photometry, due to the relatively large uncertainties in Gaia parallaxes and different atmospheric composition for some of the candidates, a significant fraction of our candidates turned out to be not as massive or as hot as expected based on a detailed model atmosphere analysis (see Section \ref{fitting method}). 
Because of this, we divide our candidates into two categories: the hot/massive sample with $M \geq 0.9~M_\odot$ and $T_{\rm eff}\geq11,000$ K (blue points in Figure \ref{fig:color mag}) and the cool/non-massive sample with one or both of the parameters outside of those boundaries (red points in Figure \ref{fig:color mag}). 
We mark an outlier in Gaia colors J1509$-$4930 with $G_{\rm BP}-G_{\rm RP}>0.4$ with a blue triangle. 

Table \ref{short phys} presents the physical parameters for our entire sample. We exclude the DC white dwarf J0640+7352 and the cool DQ white dwarf J0929$-$6050 from the hot/massive sample. In addition, we removed six targets (J0613$-$6821, J0722$-$3238, J0804$-$4757, J1015$-$4912, J1418$-$1848, and J1750$-$4558) from the sample entirely because their spectra are not consistent with white dwarfs, though they are in crowded fields, and there could be source confusion and target acquisition problems within our spectroscopic setup. J0216-6817 (DBA), J0324+6757 (DBA),  and 20 DA white dwarfs have inconsistent photometric and spectroscopic masses, where the spectroscopic masses are lower than the mass cutoff for the hot/massive subsample; we exclude these stars from the hot/massive subsample (also shown in red in Figure \ref{fig:color mag}). White dwarfs in common with \citet{jewett2024} are marked with *, cool/non-massive white dwarfs are marked with **, and the objects with low spectroscopic masses are marked with *** in Table \ref{short phys}. With these cuts, our hot/massive subsample (for the merger fraction estimate) consists of 208 white dwarfs. 

\section{Model Atmosphere Analysis}
\label{model atmo}
\subsection{The Fitting Method}
\label{fitting method}

There are two main methods for calculating the physical parameters of white dwarfs; the photometric method and the spectroscopic method. 
Starting with the photometric technique \citep{bergeron19,genest19,blouin19loweff, gentilefusillo21}, we use the available SDSS $u$, Pan-STARRS $grizy$, SkyMapper $uvgriz$ photometry along with distances from \citet{bailerjones21} to constrain the effective temperature and the solid angle $\pi (R/D)^2$ with radius $R$ and distance $D$. Given Gaia distance estimates, we can then constrain the radius of the star, and its mass using white dwarf evolutionary models. We use Gaia photometry if no other optical photometry is available. We also include the Galaxy Evolution Explorer (GALEX) NUV and FUV photometry to help constrain the atmospheric composition, but they are not used in the model fits.

We convert the observed magnitudes from these surveys into average fluxes using the appropriate zero points, and compare these with the synthetic fluxes from the model atmospheres assuming the appropriate atmospheric composition. We define  $\chi^2$ as the difference in the observed flux and model flux for all the band passes with properly weighted photometric uncertainties. We minimize this using the Levenberg-Marquardt nonlinear least-squares method \citep{press86} to get the best fitting parameters and uncertainties for each star. We account for reddening using the mean $A_V$ from \citet{gentilefusillo21} with $A_V/E(B-V)=3.1$ and assume the \citet{fitzpatrick99} reddening law. 

For the DA and DB white dwarfs in our sample, we also use the spectroscopic technique for comparison. We follow the same technique used in \citet{kilic26cooldesi}, where we fit the full spectrum to determine the physical parameters.
To break the degeneracy between the hot and cold solutions, we use the photometric temperature as the initial seed in our spectroscopic fits.
We use the CO core white dwarf evolutionary models from \citet{bedard20} with $q({\rm He})\equiv \log M_{\rm   He}/M_{\star}=10^{-2}$ and $q({\rm H})=10^{-4}$ and $10^{-10}$ for H- and He-atmosphere white dwarfs, respectively. We have a priori no knowledge of the core composition for these
massive white dwarfs. However, the assumption of an ONe core tends to lower the masses by $\approx0.04~M_\odot$ \citep{jewett2025}.

General Relativistic (GR) effects become important for the most massive white dwarfs in our sample with $M\geq1.29~M_\odot$. We use the \citet{althaus23_CO} evolutionary tracks for white dwarfs, where the Newtonian CO-core models from \citet{bedard20} indicate $M\geq1.29~M_\odot$. 
For comparison, we also use the ONe core evolutionary tracks including GR effects from \citet{althaus22_ONe}. For consistency with the rest of the sample, we consider the CO-core masses in our analysis\footnote{The only star with an extrapolated mass is J0828$-$3440 (WDJ082857.56$-$344022.70) where 
we use the \citet{althaus23_CO} models for the parameters.}.

\begin{deluxetable*}{cccccccc}
\tabletypesize{\scriptsize}
\tablecolumns{8} \tablewidth{0pt}
\label{short phys}
\caption{Physical parameters for our white dwarf sample. Objects in common with \citet{jewett2024} are marked with a * symbol. Cool/non-massive white dwarfs are marked with **, and stars with spectroscopic masses lower than 0.9 $M_\odot$ are marked with ***. 
This table is available in its entirety in machine-readable format in the online journal. A portion is shown here for guidance regarding its form and content.}
\tablehead{\colhead{Object name} & \colhead{Spectral Types} & \colhead{Composition} & \colhead{Mass (CO)} &  \colhead{Mass (ONe)} & \colhead{$T_{\rm eff}$} & \colhead{Cooling Age} & \colhead{Merger Evidence} \\
& & & ($M_\odot$) & ($M_\odot$) & (K) & (Gyr)   & }
\startdata
J0003+2603 & DA & H & $1.195_{-0.160}^{+0.067}$ & & $13796_{-1054}^{+1053}$ & $1.52_{-0.63}^{+0.30}$ & \\ 
J0011$-$5953 & DAH? & H & $1.326_{-0.012}^{+0.012}$ & $1.307_{-0.011}^{+0.011}$ & $33932_{-3014}^{+3015}$ & $0.002_{-0.0004}^{+0.001}$ & M \\ 
J0012+5708 & DAH? & H & $1.207_{-0.041}^{+0.042}$ & & $34608_{-4082}^{+4081}$ & $0.10_{-0.04}^{+0.06}$ & M\\ 
J0018+2402** & DA & H & $1.290_{-0.036}^{+0.028}$ & & $9998_{-435}^{+434}$ & $2.52_{-0.02}^{+0.28}$ & \\
J0028+0947 & DAH & H & $1.337_{-0.012}^{+0.010}$ & $1.315_{-0.012}^{+0.011}$ & $24604_{-1134}^{+1134}$ & $0.005_{-0.001}^{+0.001}$ & M \\ 
J0043+8003 & DA & H & $1.329_{-0.031}^{+0.029}$ & $1.308_{-0.026}^{+0.028}$ & $31401_{-6967}^{+6967}$ & $0.003_{-0.001}^{+0.003}$ & \\ 
J0049$-$0810 & DA & H & $1.293_{-0.009}^{+0.008}$ & & $20638_{-572}^{+572}$ & $0.008_{-0.001}^{+0.001}$ & \\ 
J0050$+$3138* & He-DA & log(H/He)=$-$3.00 & $1.215_{-0.009}^{+0.009}$ & & $12519_{-221}^{+221}$ &  $1.67_{-0.05}^{+0.05}$ &  \\
J0056$-$3347 & DAH? & H & $1.288_{-0.006}^{+0.007}$ & & $26223_{-669}^{+669}$ & $0.003_{-0.0003}^{+0.0003}$ & M \\ 
\enddata
\tablecomments{Merger Evidence: A = Unusual atmospheric composition, M = Magnetism, R = Rapid rotation, V = (large tangential) Velocity.}
\end{deluxetable*}

\subsection{DA White Dwarfs}
Out of the entire 293 targets in the sample, 155 are DA white dwarfs, including 109 that belong in the hot/massive subsample. 
For the model atmosphere analysis of DA white dwarfs, we rely on the pure H composition LTE models from \citet{tremblay09} with improved Stark broadening profiles. For DAs with temperatures hotter than 30,000 K, we include non–local thermodynamic equilibrium (NLTE) effects using TLUSTY \citep[][see also \citealt{genest14,genest19dadb}]{hubeny95}. Our model grid covers $T_{\rm eff}$ ranging from 1500 to 150,000 K and $\log{g}$ from 6.0 to 9.5. Convection is included using the ML2/$\alpha=0.7$ parameterization of the mixing-length theory. We include the 3D atmosphere model corrections from \citet{tremblay13} for all of the spectroscopic parameters presented in our figures and tables. For the cooler models, we include non-ideal effects in the equation of state \citep{blouin18}.

\begin{figure}
\centering
\includegraphics[width=3in,clip=true, trim=0.4in 0.8in 0.2in 1.1in]{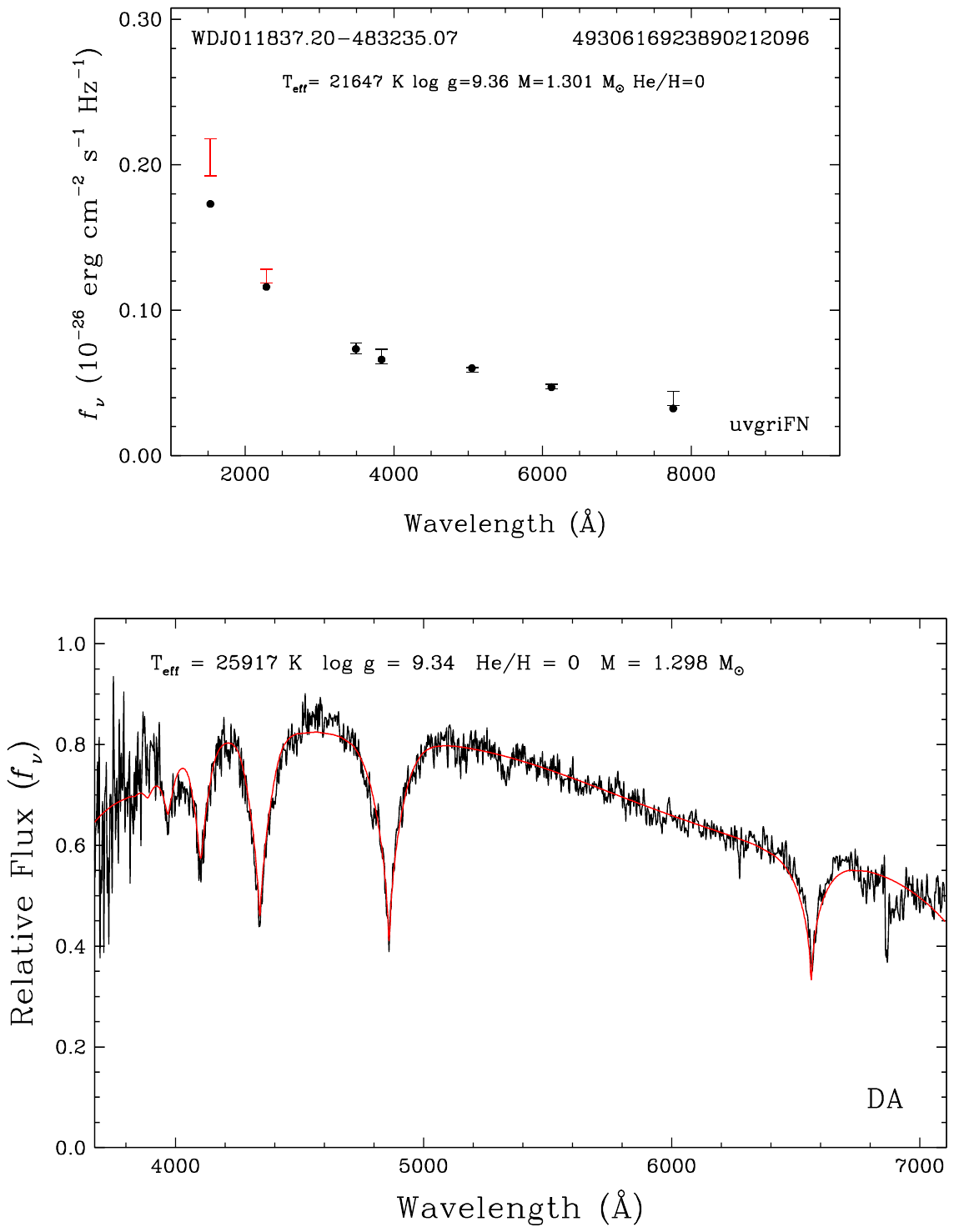}
\caption{Example model fit for a DA white dwarf. The top panel shows the best-fitting pure hydrogen (filled dots) atmosphere model to the photometry (error bars). It also includes the white dwarf name, Gaia Source ID, and the photometry used in the fitting. The bottom panel shows the spectroscopic fit (red line) to the Gemini spectrum of the same target (black line). The complete figure set is available in the online journal.}
\label{fig:da}
\end{figure}

Figure \ref{fig:da} shows our photometric (top) and spectroscopic (bottom panel) model fits to one of the DA white dwarfs in the hot/massive subsample. The top panel shows the GALEX FUV and NUV and SkyMapper $uvgri$ photometry (error bars) along with the predicted fluxes (filled dots) from the best-fitting pure hydrogen atmosphere model for J0118$-$4832 (WDJ011837.20$-$483235.07). The labels in the same panel give the white dwarf name, Gaia Source ID, and the photometry used in the fitting. Here, GALEX data is not included in the model fit (shown as red error bars), but they are used as an atmospheric composition indicator. The best-fitting model for this DA white dwarf with $T_{\rm eff}=21,647\pm1943$ K, $\log{g}=9.36$, and $M=1.301~M_\odot$ including GR corrections from \citet{althaus23_CO} provides a very good match to the GALEX data in the UV, confirming a pure H atmosphere composition. 

The bottom panel in Figure \ref{fig:da} shows the spectroscopic fit (red line) to the Gemini spectrum of the same target (black line). Excluding the atmospheric B band near 6860-6890 \AA, a pure H atmosphere model with $T_{\rm eff}=25,917$ K, $\log{g}=9.34$, and
$M=1.298~M_\odot$ provides an excellent match to the optical spectrum of this target. The internal errors in spectroscopic fits can be arbitrarily small for high signal-to-noise ratio spectra, but systematic uncertainties are significant. Using repeat observations of the same stars, \citet{liebert05} estimate external uncertainties of 0.038 dex in $\log{g}$ and 1.2\% in $T_{\rm eff}$ for DA white dwarfs. While the photometric and spectroscopic masses for J0118$-$4832 are in agreement within the errors, the spectroscopic temperature is slightly higher, but still consistent with the photometric temperature within $\approx2\sigma$. The spectroscopic method requires very well flux-calibrated spectra to fit the entire spectrum (instead of using the normalized line profiles). This is the case for the DESI spectra \citep{kilic26cooldesi, kilic26}, but not generally the case for all spectra. Hence, we only consider the photometric parameters for each white dwarf in our analysis. 

\begin{figure}
\centering
\includegraphics[width=3in,clip=true, trim=0.4in 0.8in 0.2in 1.1in]{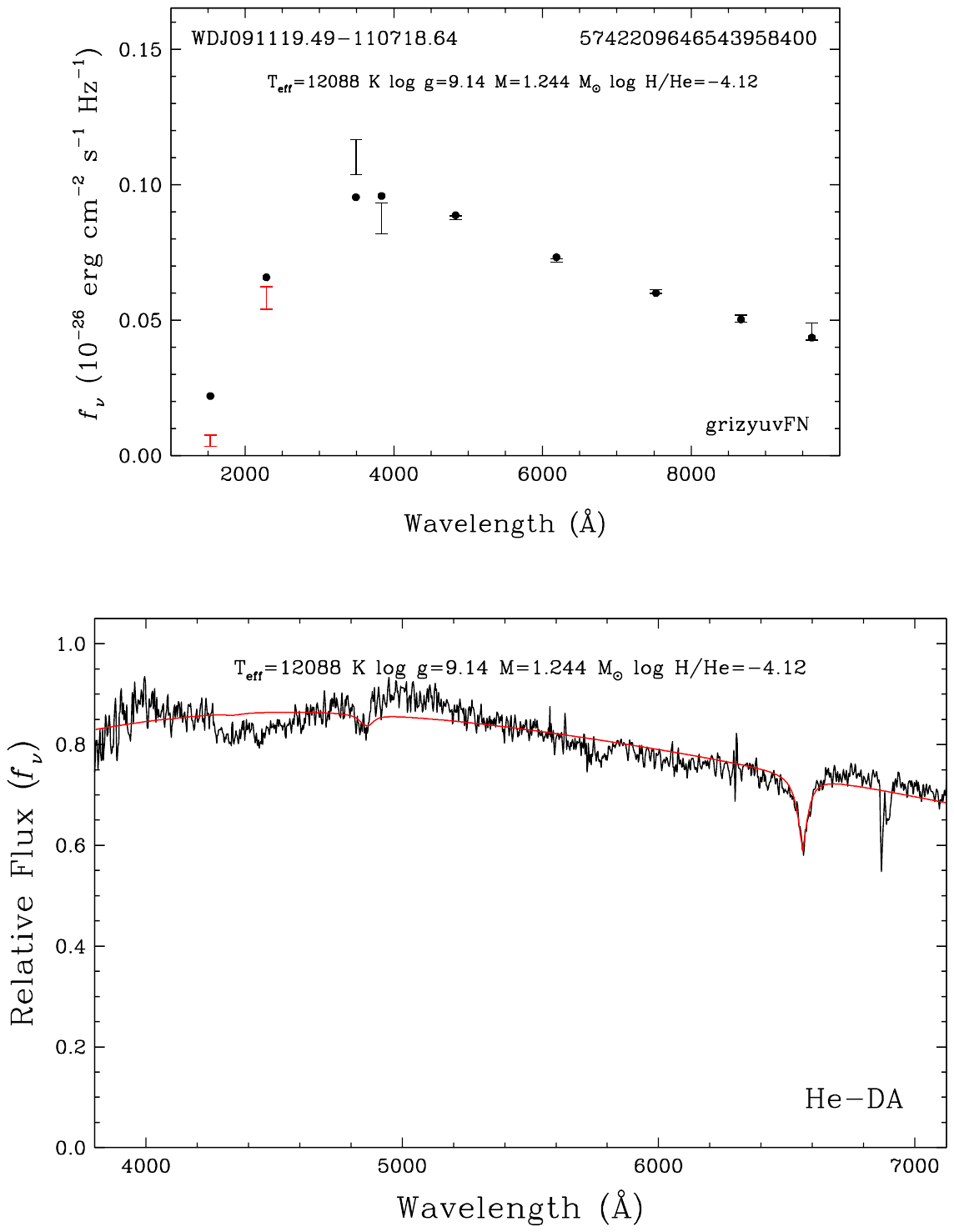}
\caption{Example model fit for a He-DA white dwarf based on a joint analysis of the photometry and spectroscopy.}
\label{fig:he-da}
\end{figure}

\subsection{He-DA White Dwarfs}

There are two DA white dwarfs in our sample where the Balmer lines are significantly weaker than expected because they are heavily broadened through van der Waals interactions in a helium-dominated atmosphere. One of these, J0050+3138, was previously confirmed as a He-DA in \citet{jewett2024}. Figure \ref{fig:he-da} shows the model fit for a new He-DA found in our study. J0911$-$1107 (WDJ091119.49$-$110718.64) shows a weak H$\alpha$ and an even weaker H$\beta$. Our mixed atmosphere model with $T_{\rm eff}=12,088$ K, $M=1.244~M_\odot$, and $\log$ H/He $=-4.12$ provides an excellent match to both the photometric spectral
energy distribution of this star and its Gemini spectrum. Even though GALEX UV photomery is not used in the fit, it also confirms the He-dominated
nature of the atmosphere. 

\subsection{DC White Dwarfs}

We find 14 DC white dwarfs with featureless spectra. J0327+2227 (WDJ032713.66+222750.67) and J0707+5612 (WDJ070753.00+561200.25) are massive and hot white dwarfs that were previously classified in \citet{jewett2024}. Because these white dwarfs are all hotter than 15,000 K, they should show spectral features from H or He. Hence, the only explanation for the missing features is the presence of a strong magnetic field that shifts and spreads out the spectral features beyond recognition,
creating a featureless spectrum. J0327+2227 and J0707+5611 also show rapid rotation with periods of 58.6 and 63 min, respectively \citep{jewett2024}. We classify these white dwarfs as likely magnetic.

J0718+3731 (WDJ071816.42+373139.24) was previously classified as a DC white dwarf in \citet{jewett2024}. Since then, new spectra have been published showing that it is not a DC white dwarf. J0718+3731 is a ``double-faced'' white dwarf with one hemisphere dominated by hydrogen and the other dominated by helium, making it a DABH white dwarf with a magnetic field strength of 8 MG \citep{cheng25J0718}. This star also has a reported rotation period of 11.27 min \citep{jewett2024}. This white dwarf is an excellent example of a merger product.

The other 12 DC white dwarfs do not meet the hot/massive criteria, and are included in our cool/non-massive subsample. Because these
DCs are cooler than 11,000 K, we cannot say whether they are featureless due to strong magnetism or simply because they have He atmospheres, 
since the helium lines disappear below $T_{\rm eff}\sim$11,000 K. For our model atmosphere analysis of cool DC white dwarfs, we use He-dominated atmosphere models
with $\log$ H/He = $-5$, since trace amounts of H are required to explain the location of the He-dominated atmosphere white dwarfs in Gaia color-magnitude diagrams \citep{bergeron19}. 

\begin{figure}
\centering
\includegraphics[width=3in,clip=true, trim=0.4in 0.8in 0.2in 1.1in]{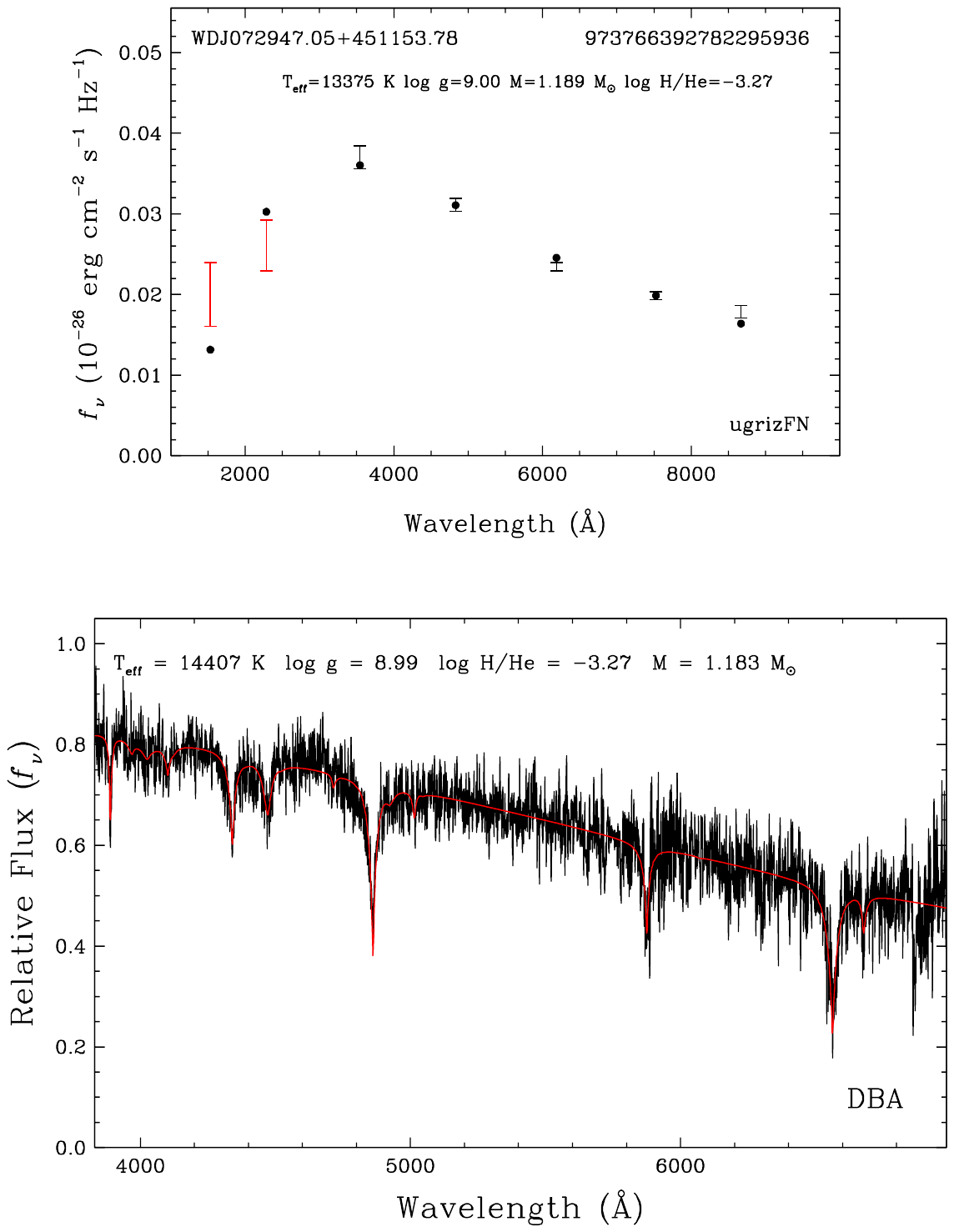}
\caption{Example model fits to a DBA white dwarf. The top panel shows the photometric fit, and the bottom panel shows that a model with
$\log$ H/He = $-3.27$ provides an excellent match to the observed MMT spectrum of this target.}
\label{fig:dba}
\end{figure}

\subsection{DB White Dwarfs}
We find four DB white dwarfs in our sample, including three DBA white dwarfs that also show H lines. Trace amounts of hydrogen are commonly seen among the DB population in the solar neighborhood \citep{koester15}. For our model atmosphere analysis, we use the hydrogen lines to constrain the H/He ratio in the atmosphere, and enforce this composition in the photometric fits. We rely on the LTE model atmospheres from \citet{bergeron11}, with improvements to neutral broadening presented in \citet{genest19} and the improved Stark-broadened \ion{He}{1} profiles from \citet[the so-called B25 semi-analytical profiles including line dissolution][]{tremblay26}. 

Only one of the DBs in our sample have consistent photometric and spectroscopic solutions that place it in the hot/massive subsample. J0955+2122 is not massive, and J0216$-$6817 and J0324+6757 do not have consistent photometric and spectroscopic masses.
Figure \ref{fig:dba} shows our model fits to J0729+4511 (WDJ072947.05+451153.78). Here a model with $M=1.189_{-0.112}^{+0.066}~M_\odot$, $T_{\rm eff}=13,375\pm841$ K, and $\log$ H/He $=-3.27$ provides an excellent match to the spectral energy distribution of this star. The spectroscopic temperature is slightly higher, but consistent with the photometric temperature within the errors. The best-fitting model also provides an excellent match to the GALEX UV data. 

Ultramassive DBs are uncommon in the solar neighborhood \citep{kilic25sdss}. \citet{jewett2024} found six DBs in their study of the massive white dwarfs in the 100 pc sample. However, all six are either strongly magnetic or show evidence of rapid rotation, including a rapidly rotating DBA \citep{pshirkov20}.
Hence, given its high mass and helium-dominated atmosphere, J0729+4511 is likely a merger remnant. 

\subsection{DQ White Dwarfs}

\begin{figure}
\centering
\includegraphics[width=3in,clip=true, trim=0.4in 0.8in 0.2in 1.1in]{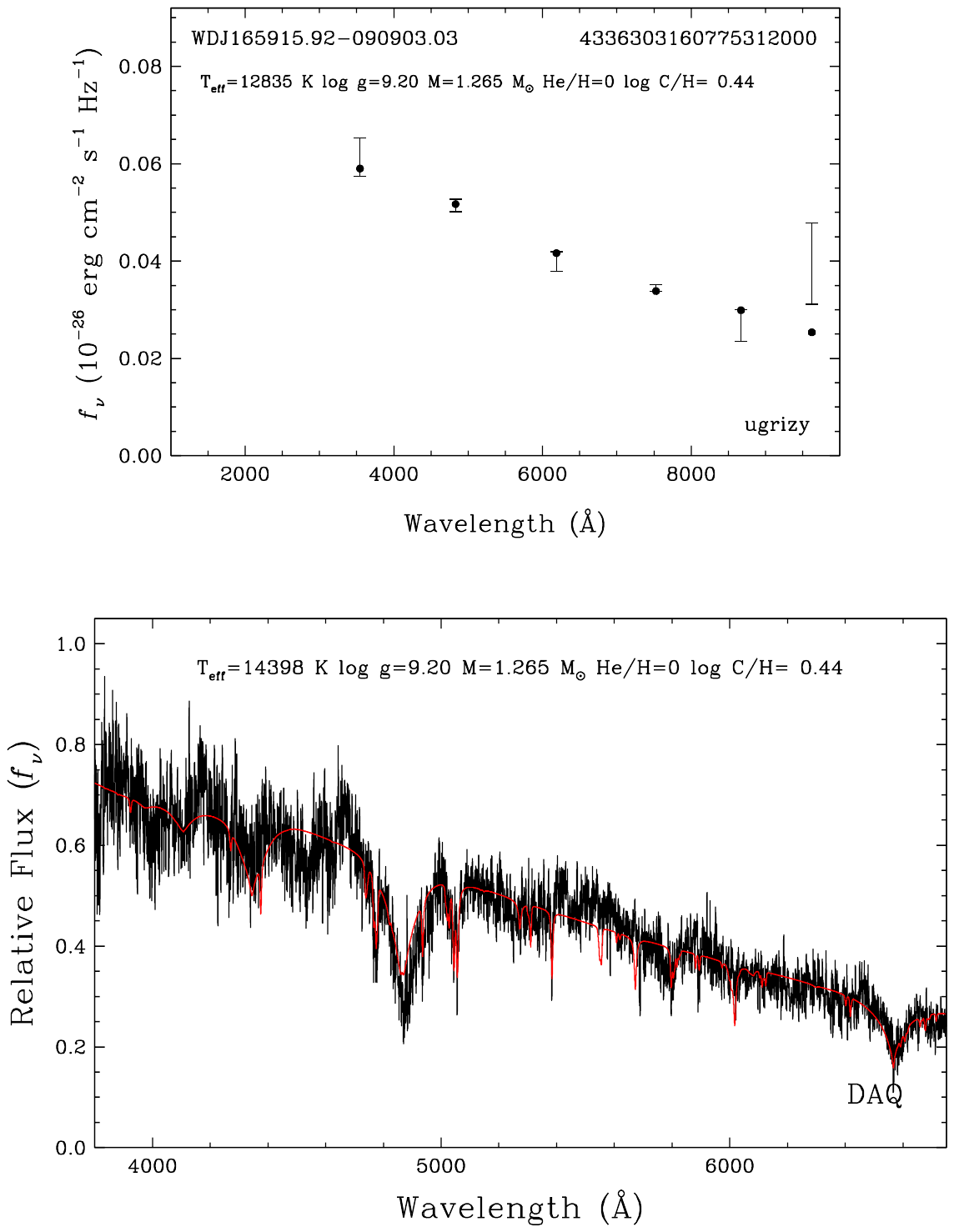}
\caption{Model fit for a newly identified DAQ white dwarf, J1659$-$0909. The top panel shows the photometric fit and the bottom panel shows the spectroscopic fit.}
\label{fig:daq}
\end{figure}

\begin{figure*}
\centering
\includegraphics[width=3in,clip=true, trim=0.4in 0.8in 0.2in 1.1in]{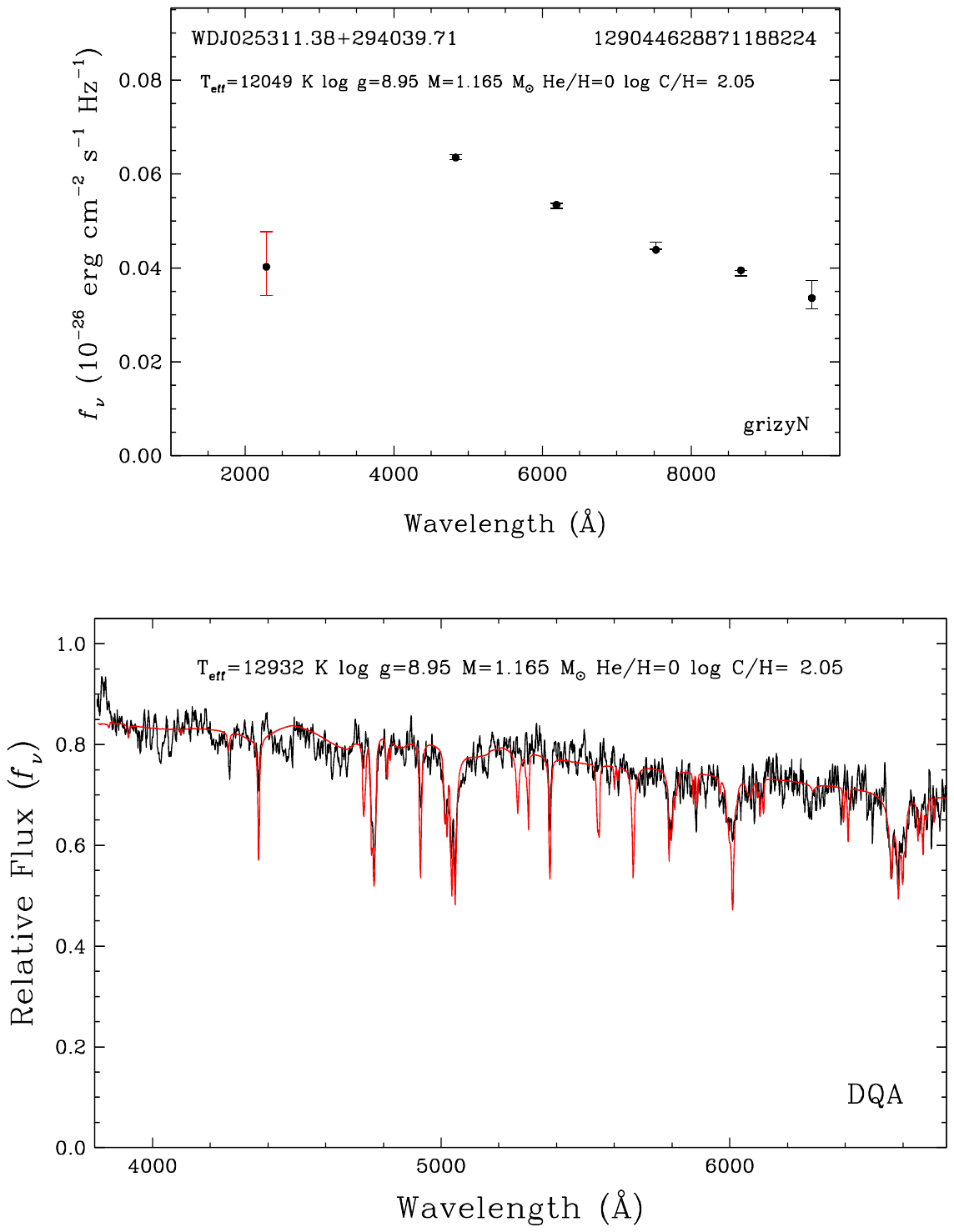}
\includegraphics[width=3in,clip=true, trim=0.4in 0.8in 0.2in 1.1in]{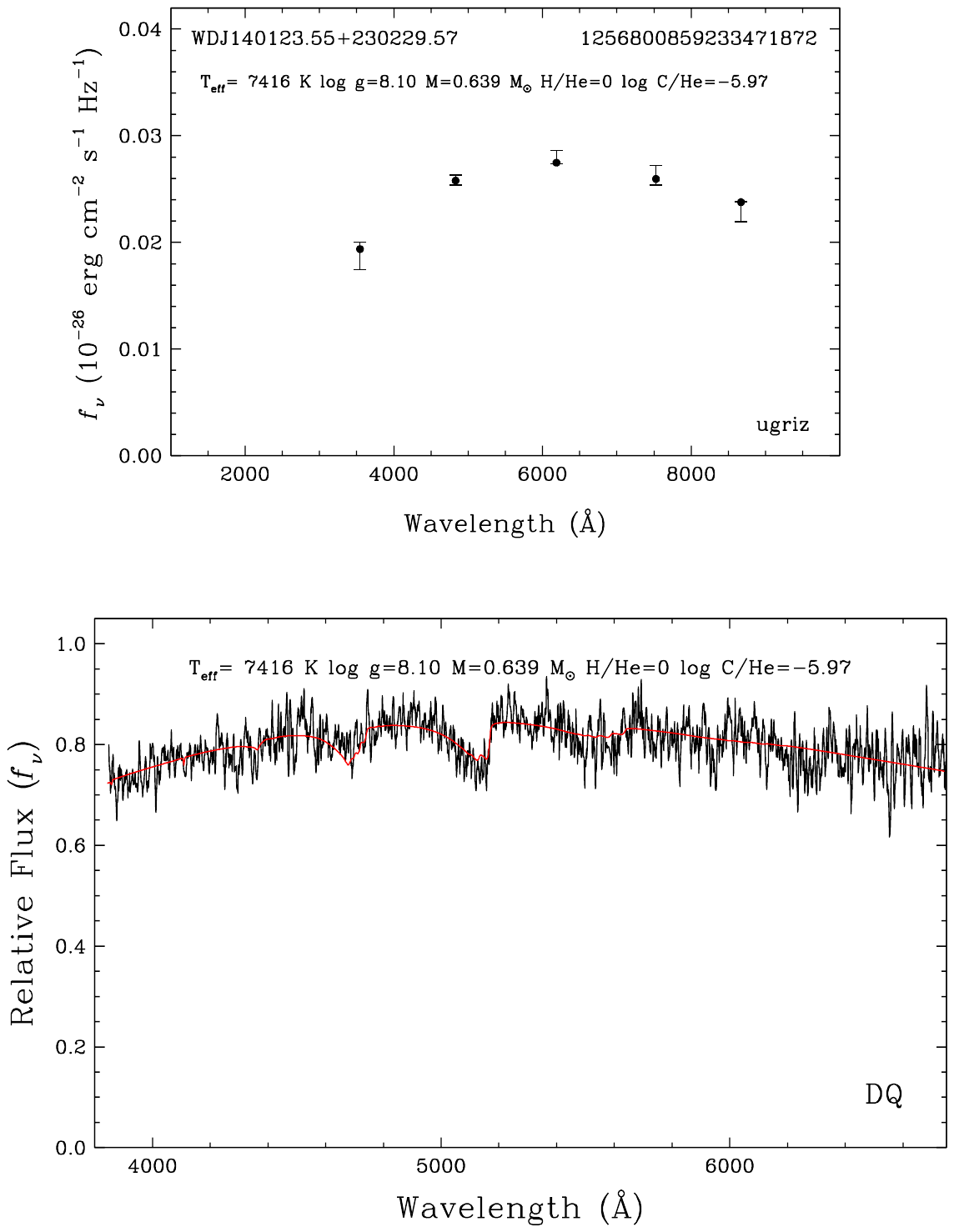}
\caption{Example model fits to a warm DQA (left panels) and a cool DQ (right panels). The top panels show the photometric fits and the bottom panels shows the spectroscopic fits.} 
\label{fig:dqs}
\end{figure*}

DQ white dwarfs can be classified into two main types: cool DQs and warm DQs. Cool DQs typically have effective temperatures below 10,000 K, show molecular Swan bands in their spectra, and have atmospheres dominated by helium with trace amounts of carbon. The trace carbon is consistent with convective dredge up from the interior of the star \citep{pelletier86, bedard22}. 

Warm DQs on the other hand are very different from the cool DQs, and are very similar to the so called hot DQs. These have effective temperatures exceeding 11,000 K, tend to have high mass ($M\gtrsim1~M_\odot$, carbon dominated atmospheres, and they often show signatures of being the product of a merger event. Hot DQs often show strong magnetic fields, rapid rotation, and unusual kinematics \citep{dufour08, coutu19, koester19, kawka23}. While warm DQs are very similar to the hot DQs, they do not show a high frequency of magnetism \citep{coutu19,ouldrouis26,kilic26}. Because the temperature cut off between the warm and hot DQs is arbitrary, we refer to all of our hot/warm DQs as warm DQs. More importantly, the majority of the warm DQs are found on the crystallization sequence \citep{kilic24daq,jewett2024,kilic25}, indicating that they suffer from relatively long cooling delays due to $^{22}$Ne distillation \citep{blouin21,bedard24}. Because of these characteristics, warm DQ white dwarfs are the best examples of merger remnants \citep{dunlap15supernova}.    

Many warm DQs show evidence of hydrogen. The most H-rich ones, DAQ white dwarfs, show hydrogen features that are stronger than carbon lines.
The number of DAQ white dwarfs has increased dramatically since the prototype \citep{hollands20}. Through targeted surveys of massive white dwarfs \citep{jewett2024,kilic24daq}, UV-faint objects \citep{kilic25}, and Q-branch white dwarfs \citep{ouldrouis26}, the number of DAQs increased to 17.
\citet{kilic26} recently reported the discovery of 6 new DAQs in the Dark Energy Spectroscopic Instrument Data Release 1. 
Here, we report the discovery of yet another DAQ, J1659$-$0909 (WDJ165915.92$-$090903.03), bringing the total number of DAQs known to 24. 

\begin{figure*}
\centering
\includegraphics[width=3in,clip=true, trim=0.4in 0.8in 0.2in 1.1in]{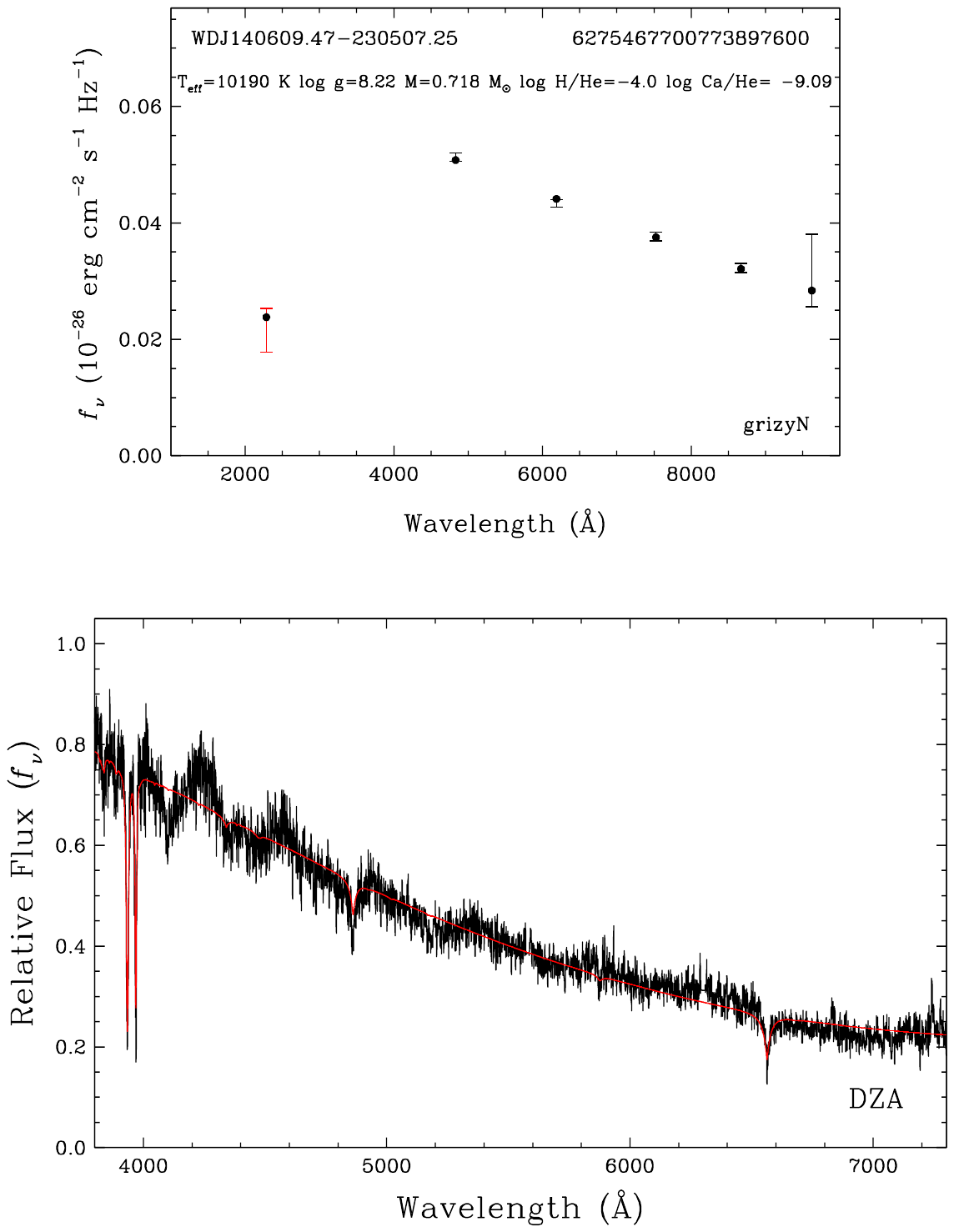}
\includegraphics[width=3in,clip=true, trim=0.4in 0.8in 0.2in 1.1in]{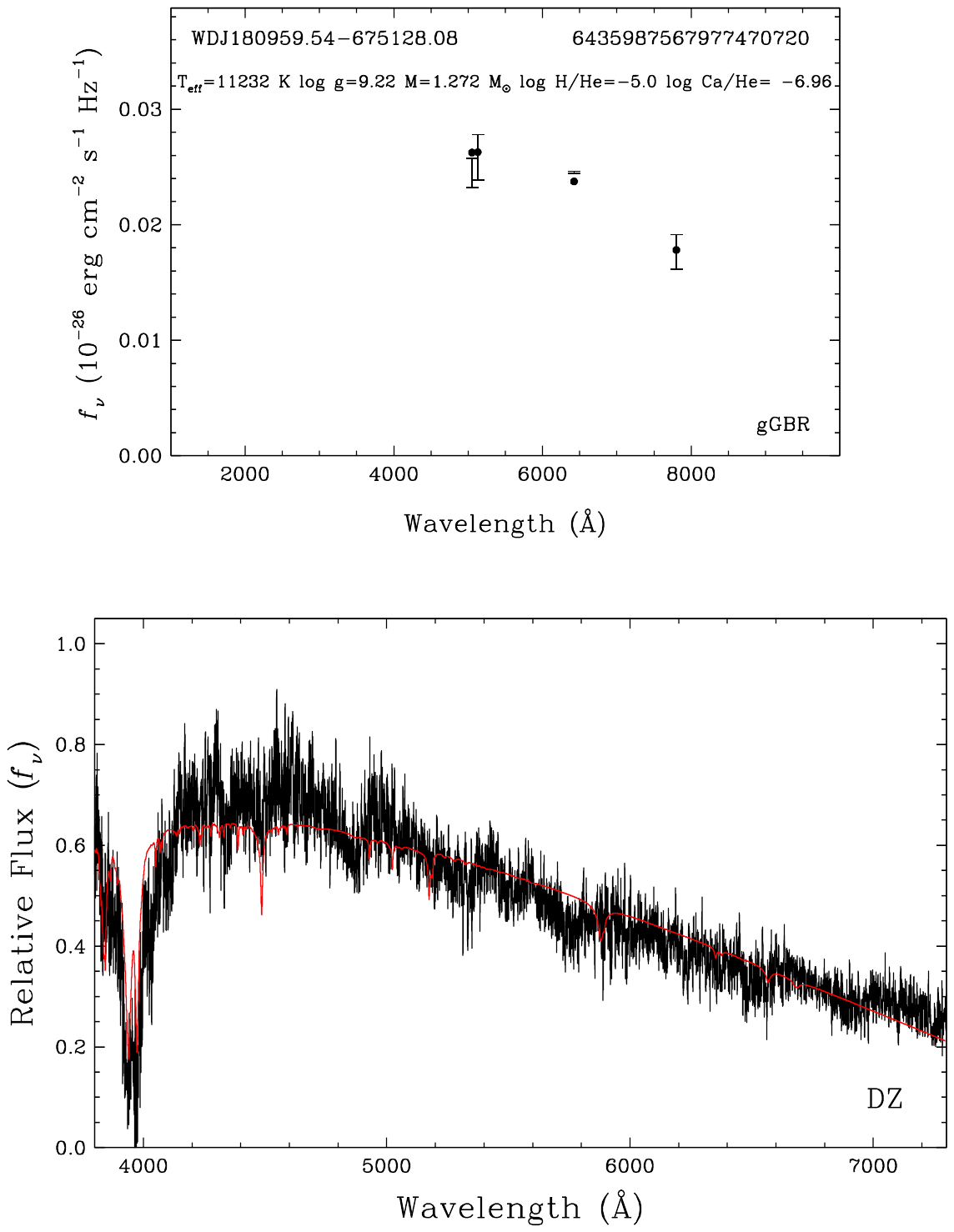}
\caption{Model fits for two DZ white dwarfs: J1406$-$2305 and J1809$-$6751. J1406$-$2305 shows strong Ca lines as well as H$\alpha$ and H$\beta$, whereas J1809$-$6751 only shows strong Ca lines.}
\label{fig:dz}
\end{figure*}

Figure \ref{fig:daq} shows our model fits to J1659$-$0909. We rely on the He-free model atmosphere grid from \citet{kilic25}. 
We use the photometric spectral energy distribution (top panel) to constrain $T_{\rm eff}$ and $\log{g}$ (i.e., the radius), and use the spectrum
(bottom panel) to determine the C/H ratio in the atmosphere. We iterate between the spectroscopic and photometric fits until a consistent solution is
found. This method provides joint constraints on the masses and C/H ratios in the atmosphere, and independent estimates of the effective temperature
from photometry and spectroscopy. Here, a model with $M=1.265_{-0.040}^{+0.036}~M_\odot$ and $\log$ C/H = 0.44 provides an excellent match to the
observed photometry and Magellan spectrum of this target. We find an issue with modeling another DQA, J1853$-$2515 (WDJ185350.12$-$251549.90). When we model this white dwarf using the same He-free model for the other DQAs we find that we do not get a good fit to the spectrum. When we model this white dwarf using the He rich solution, we get a much better fit to the spectrum. We classify this star as a DQA, but in order to constrain the physical parameters of this star more accurately we need a higher S/N spectrum as the one we currently have is quite noisy.  

Our sample has a total of 25 warm DQs (5 DQs, 19 DQAs, and 1 DAQ) and 2 cool DQs. Figure \ref{fig:dqs} shows example model fits for a warm DQA (left panels) and a cool DQ (right panels). For warm DQs, we rely on the same fitting method as for the DAQs and also use the He-free models. As discussed
in \citet{kilic25}, due to the lack of spectral features from helium, it is impossible to constrain the amount of helium in warm DQs, and the fits
using He-free atmosphere models provide fits just as good as the ones with helium. The resulting physical parameters (effective temperature and mass)
are also consistent between the He-free and He-rich warm DQ models. The left panels in Figure \ref{fig:dqs} show that a model with $M=1.165_{-0.029}^{+0.031}~M_\odot$, $T_{\rm eff}=12,049\pm90$ K, and $\log$ C/H = 2.05 provides an excellent match to the photometry and the MMT spectrum of the DQA J0253+2940 (WDJ025311.38+294039.71). 

For the two cool DQs in our sample, we begin with the photometric method to estimate $T_{\rm eff}$ and $\log{g}$, and then force these parameters to fit the C$_2$ Swan bands to constrain the C/He ratio. We then use this abundance in the photometric fit, and iterate between the photometric and spectroscopic
fit until a consistent solution is found. Our analysis relies on an extended version of the model grids from \citet{blouin19specevo}. The right panels in
Figure \ref{fig:dqs} shows our model fits to the cool DQ J1401+2302 (WDJ140123.55+230229.57), where a model with $M = 0.639_{-0.271}^{+0.249}~M_\odot$, $T_{\rm eff}=7416_{-197}^{+197}$ K, and $\log$ C/He $=-5.97$ provides an excellent match to the photometry and spectroscopy of this object.

\subsection{DZ White Dwarfs}

Our spectroscopic survey reveals 5 DZs, including 2 DZAs that also show H lines. We use a similar approach to our analysis of the DQ white dwarfs and
start with the photometric technique to constrain the temperature and surface gravity. We then fit the observed spectrum to constrain Ca/He with the assumption that the abundance ratios of the other heavy elements match the CI chondrites. We use the models with that abundance to update the photometric fits, and iterate between the photometric and spectroscopic fits until a consistent solution is found. 

Figure \ref{fig:dz} presents example model fits to a DZA (left panels) and a DZ white dwarf (right panels). Here, both spectra are from the optical echelle
spectrograph Mage on Magellan. Unfortunately, these two spectra suffer from some flux calibration problems with stitching the echelle orders together,
which leads to a wavy continuum in some wavelength ranges. Nevertheless, the \ion{Ca}{2} H+K lines are clearly visible and dominate these spectra in the
blue. Our best-fit model for the DZA shown in the left panels, J1406$-$2305 (WDJ140609.47$-$230507.25), with $M=0.718_{-0.189}^{+0.138}~M_\odot$, $T_{\rm eff}=10,190\pm434$ K, $\log$ H/He= $-4.0$, and $\log$ Ca/He = $-9.09$ provides an excellent match to the Ca and H lines in this object. The DZ shown in the right panels, J1809$-$6751 (WDJ180959.54$-$675128.08), has even stronger Ca lines, which require a significantly larger Ca abundance of $\log$ Ca/He = $-6.96$. J1809$-$6751 is also massive with
$M=1.272_{-0.105}^{+0.129}~M_\odot$, and is one of the two DZ white dwarfs in our sample that are hot and massive enough to make it into the hot/massive subsample. The second DZ in the hot/massive subsample is J0634$-$4811 (WDJ063434.64$-$481121.84) with log(Ca/He)=$-$7.43, $M=1.185_{-0.178}^{+0.224}~M_\odot$, and $T_{\rm eff} = 9506_{-2967}^{+2967}$ K. The effective temperature is cooler than 11,000 K, but the $1\sigma$ error places it within our selection criteria, and therefore it is included in the hot/massive subsample.

\begin{figure*}
\includegraphics[width=2.5in, clip=true, trim=0.4in 0.8in 0.2in 1.1in]{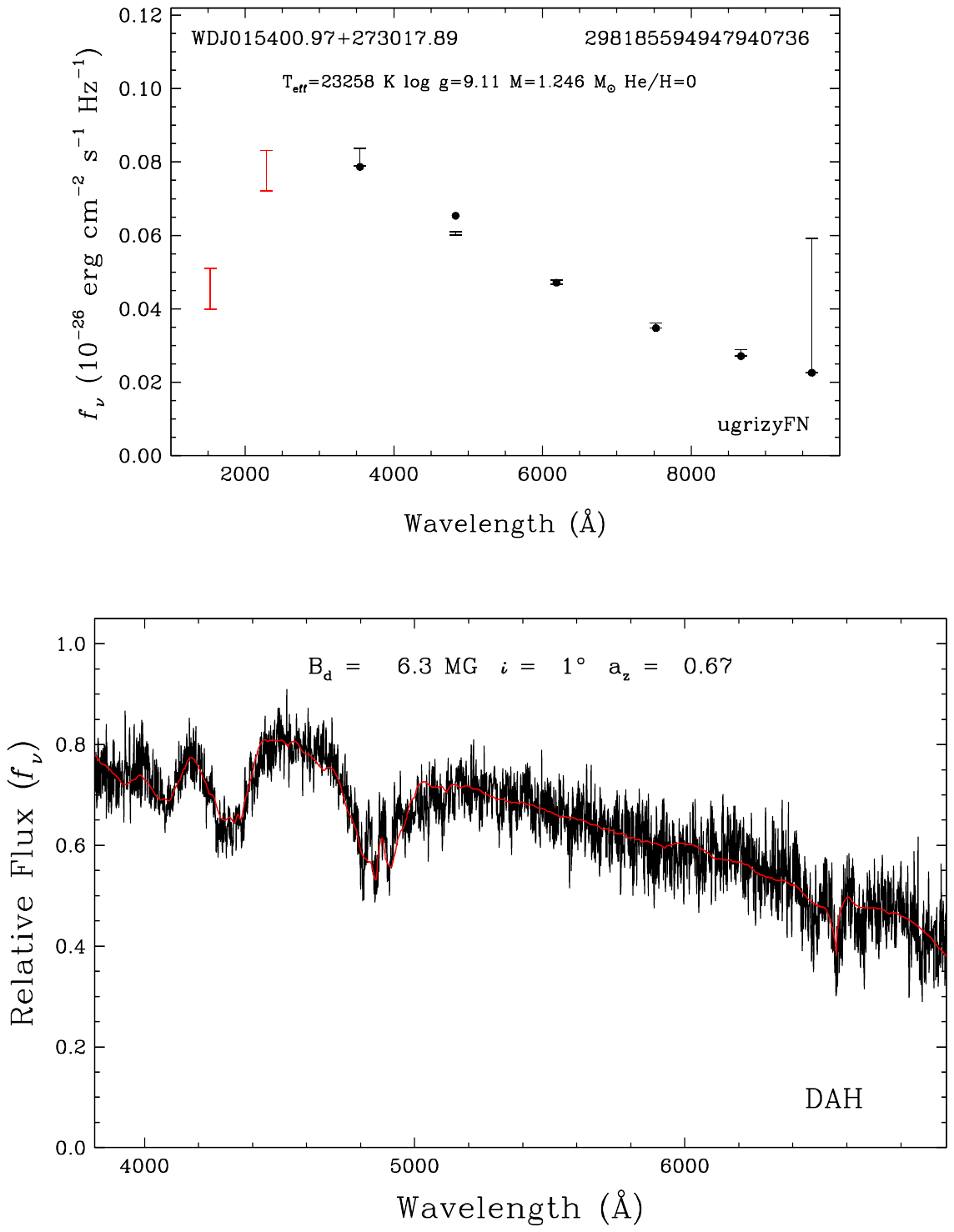}
\includegraphics[width=2.5in, clip=true, trim=0.4in 0.8in 0.2in 1.1in]{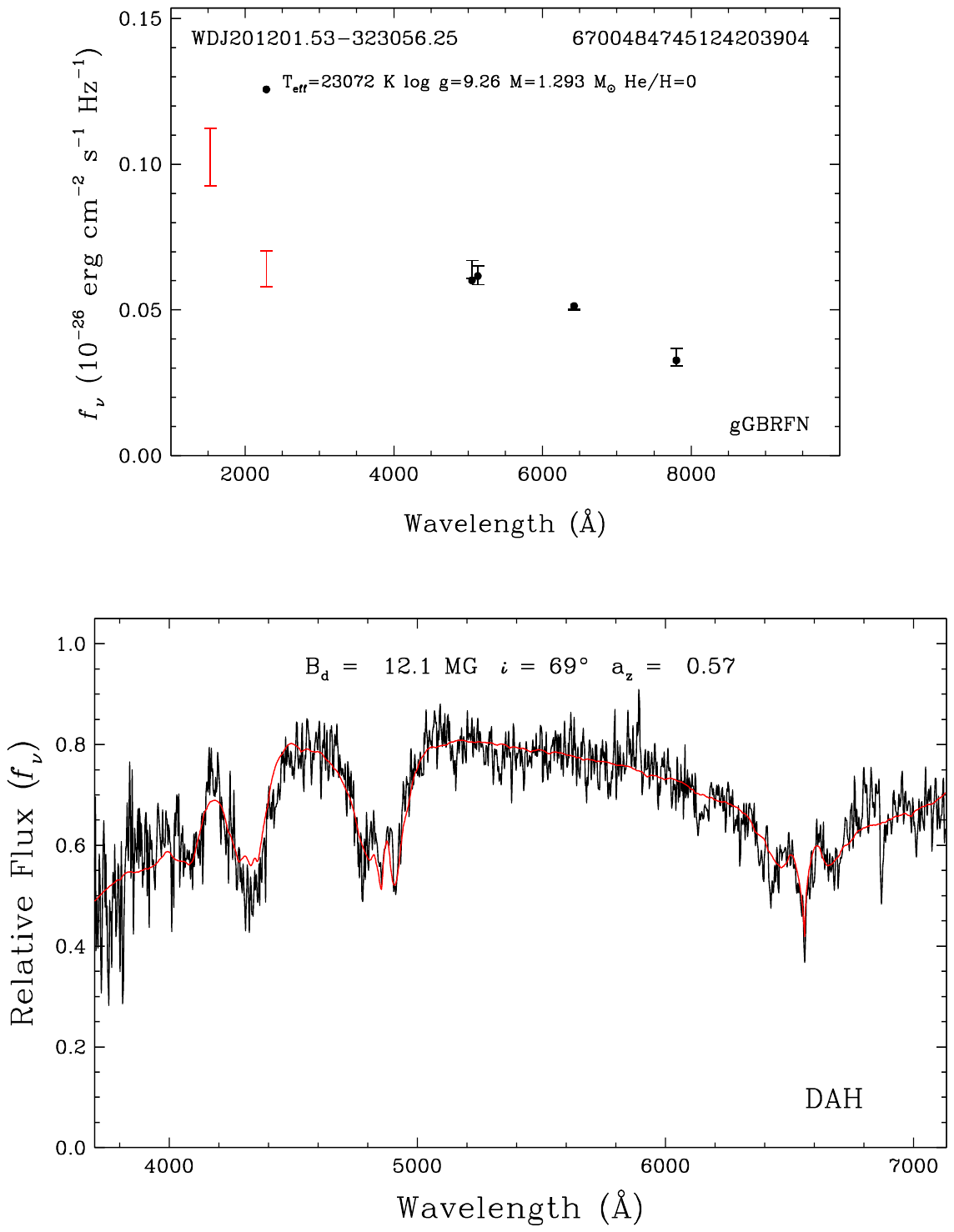}
\includegraphics[width=2.5in, clip=true, trim=0.4in 0.8in 0.2in 1.1in]{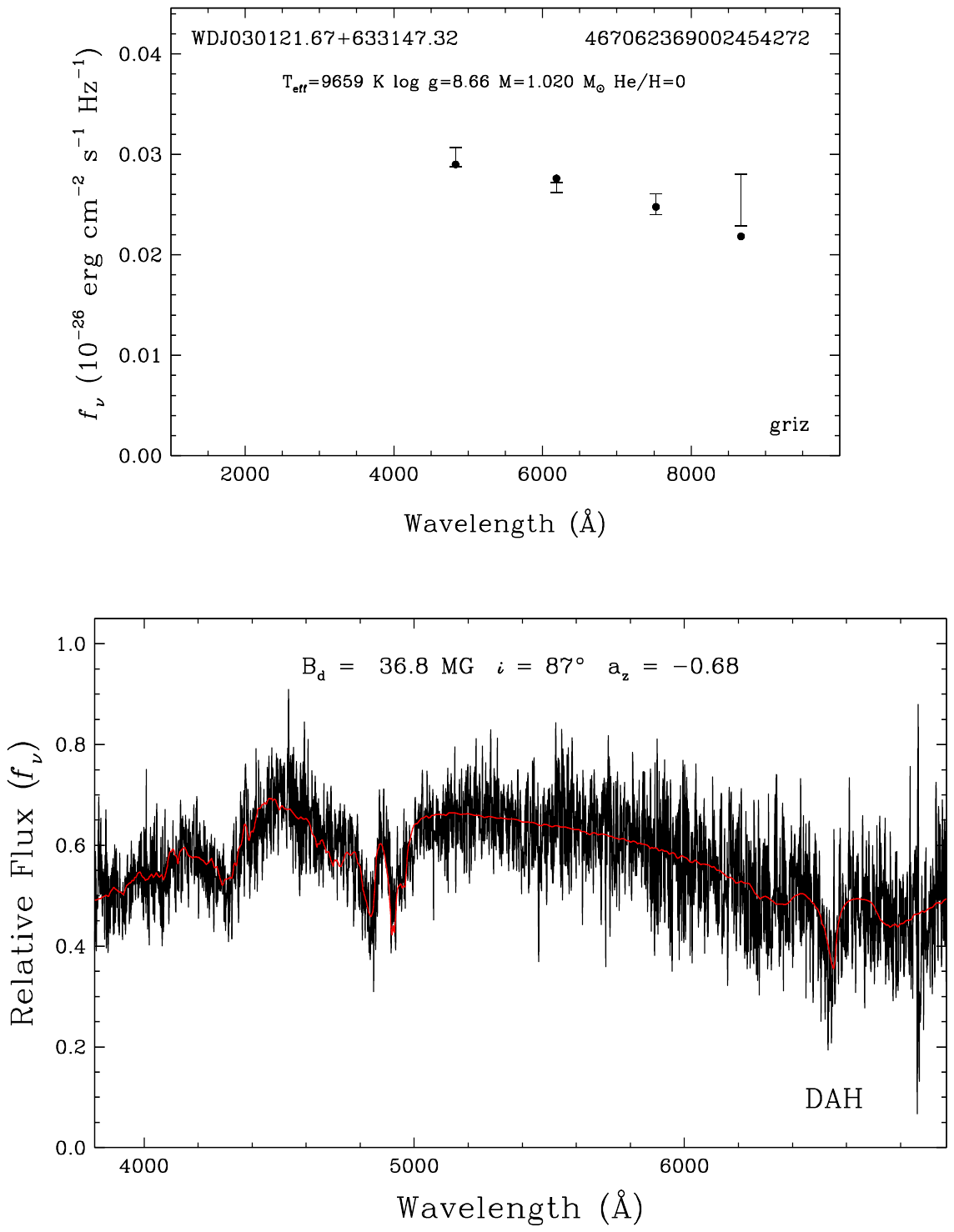}
\caption{Example model fits for 3 magnetic white dwarfs. Top panels shows the photometric fit, and the bottom panel shows the spectroscopic fit and magnetic field parameters. Each of these white dwarfs clearly show Zeeman split features. J0154+2730 (left panels) likely has a patchy atmosphere.} 
\label{fig:magfits}
\end{figure*}

The discovery of these two massive DZs is significant as it is rare to find DZs at high masses. \citet{mccleery20} find a significant difference in the mean mass between DA and DAZ white dwarfs in the 40 pc sample. 
Based on a Hubble Space Telescope UV spectroscopy program, \citet{ouldrouis24} found that
the fraction of metal-polluted white dwarfs significantly decreases for massive white dwarfs with $M>0.8~M_\odot$. Their sample contains no polluted
white dwarfs above $1~M_\odot$. \citet{cunningham25} presents an upper mass limit of $0.72~M_\odot$ for DZ white dwarfs based on the H-dominated DZA and DAZs in the 40 pc sample. Recently, \citet{kilic26,kilic26cooldesi} presented a detailed model atmosphere analysis of the white dwarf candidates in the Dark Energy Spectroscopic Instrument (DESI) Data Release 1. Out of the 1102 metal-polluted white dwarfs in the DESI DR1 sample, only two,
J1414+5438 and J1743+2512, have masses likely above a solar mass, though they have relatively large errors with $M=1.053\pm0.166~M_\odot$ and
$M=1.064\pm0.101~M_\odot$, respectively. Searching the MWDD also reveals only 27 DZ white dwarfs with mass greater than $1~M_\odot$. All but two of these are cool DZs. The two DZs with $T_{\rm eff}>11,000$ K are J1036+4837 (Gaia DR3 834176141731355392) and J1104+0711 (Gaia DR3 3818299730867320704), though both stars have a large discrepancy between the photometric and spectroscopic masses \citep{genest19}. Hence, massive DZ white dwarfs are uncommon. Nevertheless, the presence of the ultramassive DZs J1809$-$6751 and J0634$-$4811 in our sample with $M=1.272_{-0.105}^{+0.129}~M_\odot$ and $M=1.185_{-0.178}^{+0.224}~M_\odot$ demonstrates that metal pollution can still occur in ultramassive white dwarfs,
albeit it is very rare and these mass measurements have relatively large errors. More extensive statistical analysis on a larger sample of metal-polluted white dwarfs would be helpful.

\subsection{Magnetic White Dwarfs}

Our sample contains a total of 73 confirmed or suspected magnetic white dwarfs, with 67 of these part of the hot/massive subsample. Excluding the likely magnetic hot DC white dwarfs discussed previously, we classify these magnetic white dwarfs as DAH\footnote{J0118$-$0156 (WDJ011810.31-015612.23) was classified as a DAH in \citet{jewett2024} and has been reclassified as a DQH? in \citet{kilic26}. We use the spectrum from \citet{jewett2024} but update the parameters to include GR effects.}. To model these stars, we follow the approach in \citet{kilic26}. We create a grid of specific intensities for each star. At the surface, the specific intensities defined as $I(\nu,\mu,\tau_{\nu}=0)$ are found solving the radiative transfer equation for different field strengths and $\mu$, where $\mu= \cos \theta$. Here, $\theta$ is defined as the angle between the angle of propagation of light and the normal to the surface of the star. Total line opacity is determined as the sum of the individual Stark-broadened Zeeman components, and line displacements and oscillator strengths of the Zeeman components of H$\alpha$ through H$\delta$ and the neutral He lines kindly provided to us by S.~Jordan \citep[see also][]{hardy23,moss24}. Then, the total line opacity is normalized to that resulting from the zero-field solution. During the fitting procedure, the Eddington flux H$_\nu$ is calculated by numerically integrating the specific intensity across the visible surface of the disk for specified values of dipole strength ($B_d$), dipole offset ($a_z$), and viewing angle ($i$). We use the {\tt PIKAIA} genetic algorithm \citep{charbonneau95} to find the best-fitting magnetic model.


Figure \ref{fig:magfits} shows example fits for three magnetic DA white dwarfs with various field strengths. J0154+2730 (WDJ015400.97+273017.89, left) shows a clearly split H$\beta$ feature, and we find a magnetic field strength of 6.3 MG under the assumption of a homogeneous, pure H atmosphere. However, the split H$\beta$ components are significantly sharper than predicted by the model, which likely indicates a patchy atmosphere where the magnetic field is restricted to a small portion of the surface \citep{moss24}. 
J2012$-$3230 (WDJ201201.53$-$323056.25) and J0301+6331 (WDJ030121.67+633147.32) (middle and right) have higher magnetic field strengths of 12.1 MG and 36.8 MG, respectively. The strongest fields we observe in our sample are for J0951$-$1517 (WDJ095150.85$-$151712.26) and J2255+0710 (WDJ225513.47+071001.05), which have reported field strengths of 382 MG and 230 MG \citep{jewett2024}. Table \ref{tab:mags} presents the results from our offset dipole model fits. Here, we only include the fits where we are confident in the magnetic fit, as some of the highly magnetic objects do not show clear Zeeman split features, making the fitting process more complicated.

\begin{deluxetable*}{ccrrr}
\tabletypesize{\scriptsize}
\tablecolumns{5} \tablewidth{0pt}
\tablecaption{Results from our offset dipole fits for a selection of magnetic white dwarfs in our sample.} \label{tab:mags}
\tablehead{\colhead{Object name} & \colhead{Spectral Type} & \colhead{$B_d$} &  \colhead{i} & \colhead{$a_z$}\\
& & (MG) & ($^{\circ}$) & ($R_{\star}$) }
\startdata 
J0154+2730 & DAH &6.7 & 1 & 0.67 \\ 
J0301+6331 & DAH &36.8 & 87 &$-$0.68 \\ 
J0318+7923 & DAH &15.5 & 53 &0.56 \\ 
J0518$-$4831 & DAH &12.6 & 25 &0.70 \\ 
J0546+2422 & DAH &7.1 & 18 &0.65 \\ 
J0633$-$5909 & DAH &16.6 & 44 &0.67 \\ 
J0718+3731 & DABH & 8 & \nodata & \nodata \\
J0733$-$1704 & DAH &5.7 & 45 &0.70 \\ 
J0951$-$1517 & DAH  & 382  & 45 & 0.30 \\
J1154$-$5758 & DAH &51 & 71 &0.26 \\ 
J1324$-$6109 & DAH &45.6 & 58 &0.51 \\ 
J1352$-$0555 & DAH &11.1 & 53 &0.59 \\ 
J1544$-$5125 & DAH &8.8 & 0 &$-$0.68 \\ 
J1727+2351 & DAH & 8.9 & 19 &0.70 \\ 
J1732$-$6005 & DAH &51.4 & 75 &0.25 \\ 
J1824$-$4732 & DAH &9.6 & 19 &0.67 \\ 
J1931$-$1431 & DAH &4.3 & 11 &0.70 \\ 
J2012$-$3230 & DAH &12.1 & 69 &0.57 \\ 
J2255+0710 & DAH & 230 & 0 & 0.21 \\
J2256+4651 & DAH &6.7 & 22 &0.69 \\ 
J2257$-$3550 & DAH &1 & 9 &0.70 \\ 
J2322+3944 & DAH &9.3 & 24 &0.70 \\ 
J2335+2912 & DAH &26.7 & 68 &0.38 \\ 
\enddata
\tablecomments{Values for J0718+3731 are adopted from \citep{cheng25J0718}. J0951$-$1517 and J2255+0710 are from  \cite{jewett2024}.}
\end{deluxetable*}  

\section{The Sample of Hot and Massive White Dwarfs}
\label{mass dis}
Though our initial search was for white dwarfs with $M \geq 1.3~M_\odot$, we find many white dwarfs to have lower mass than this. In the case of the non-DAs, the masses we searched in \citet{gentilefusillo21} assumed a pure hydrogen atmosphere, so it is not surprising that we see such a difference in these masses as they are not pure hydrogen atmospheres. 

Another reason is that the mass uncertainties can be quite large for the masses from \citet{gentilefusillo21}, as well as our photometric mass estimates, due to the relatively large errors in Gaia parallaxes of some of these stars. The average mass error for the $M\geq1.3~M_\odot$, $T_{\rm eff}\geq11,000$ K sample from \citet{gentilefusillo21} is 0.36 $M_\odot$, where the objects with the largest mass error tend to fall into our cool/non-massive subsample.
For example, J2044+4324 (WDJ204437.92+432417.19), the \citet{gentilefusillo21} mass is $M=1.368\pm1.053 M_\odot$, whereas we find $M=0.701_{-0.230}^{+0.158}~M_\odot$ for the same star, so its parameters fall outside of our hot/massive subsample. Another example is J1551$-$6050 (WDJ155146.65$-$605033.10), where the \citet{gentilefusillo21} mass is $M=1.301\pm1.202~M_\odot$, and our photometric mass estimate is $M=0.589_{-0.319}^{+0.736}~M_\odot$, which is in agreement with the spectroscopic mass of $M=0.552~M_\odot$. 

\begin{figure}
\centering
\includegraphics[width=3.4in, clip, trim={0in 0in 0.4in 0.4in}]{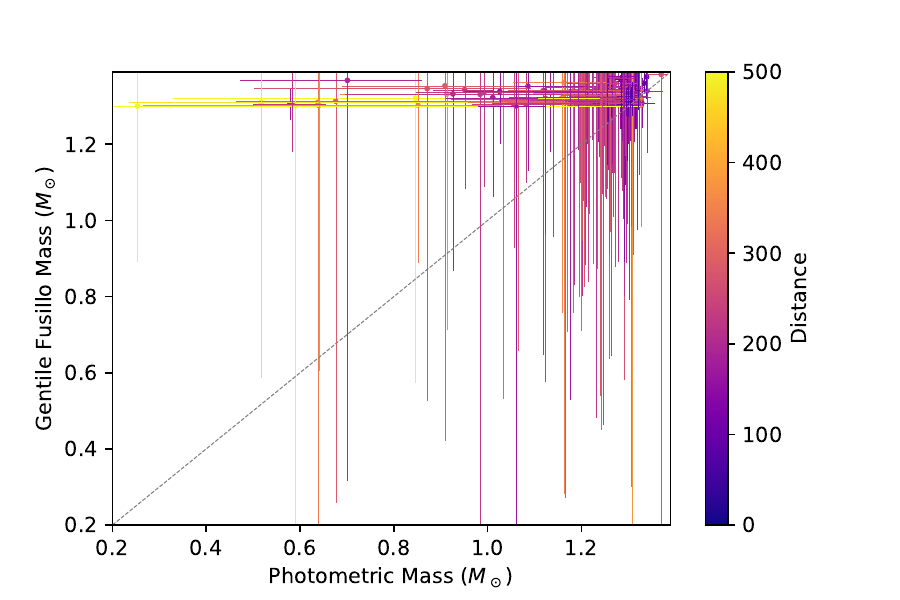}
\caption{Comparison between the photometric mass from this study versus \citet{gentilefusillo21} for the DA white dwarfs. The color of the points indicates the \citet{bailerjones21} distance and the dashed line is the one-to-one line.}
\label{fig:mass compare}
\end{figure}

Figure \ref{fig:mass compare} shows a comparison between our photometric masses and those of \citet{gentilefusillo21} for the DA white dwarfs in our sample. Each data point is color coded with the \citet{bailerjones21} distance. Generally, we see a clustering of the closest white dwarfs in the upper right hand corner, where we would expect the targets of this massive sample to be. It seems that the lower mass objects we find also tend to be more distant. \citet{gentilefusillo21} relied solely on Gaia photometry to have a uniform analysis of all white dwarfs in Gaia DR3. However, Gaia $G$,
$G_{\rm BP}$, and $G_{\rm RP}$ are not independent, as they are derived from the same observations. \citet{bergeron19} compare the photometric systems of SDSS, Pan-STARRS and Gaia in order to determine which gives the most consistent masses and temperatures. They find that combining SDSS $u$ and Pan-STARRS $grizy$ photometry generally yields the best fits and has the lowest 1$\sigma$ dispersion for mass compared to all of the photometric systems they test. They also find that by not including $u$-band photometry, especially for hot white dwarfs, the effective temperature and mass can be underestimated. Therefore, we use all available photometry for each star to get the most accurate parameters possible.
The large parallax uncertainties still limit the precision of the parameters for some of our targets. The average errors in our photometric mass measurements
are $\sim0.06~M_\odot$ for the hot/massive subsample. 

There are similar issues with the temperature measurements, too. For example, J0212+2224 (WDJ021202.85+222419.51) has $T_{\rm eff}=19,643\pm9024$ K in the \citet{gentilefusillo21} catalog. This is a large error, potentially placing the white dwarf outside of the temperature criteria we set. We find $T_{\rm eff}=8756 \pm247$ K, so this white dwarf falls outside of the parameters for our hot/massive subsample. 
Generally we find the mass/temperature estimates from \citet{gentilefusillo21} are consistent with our measurements within the errors. However, with the additional
photometry and the spectra used in our analysis, we obtain smaller errors in mass and temperature. 

The mass and temperature discrepancies affect 50 white dwarfs from our initial sample, including 26 DA, 6 magnetic, 12 DC, 2 DQ, 1 DB, 1 DZ, and 2 DZA white dwarfs. Additionally, we exclude 22 stars (20 DA and 2 DBA) where the spectroscopic mass is too low to be in our subsample and is in disagreement with the photometric mass. Excluding these stars, we are left with 208 massive and hot white dwarfs: 109 of these are DA white dwarfs, 2 are He-DA, 2 DC, 1 DBA, 1 DAQ, 19 DQA, 5 DQ, 2 DZ, and 67 are magnetic white dwarfs. The physical parameters for all of the objects in our sample are presented in
Table \ref{short phys}. 

\section{The Search for Merger Signatures}
\label{search for merger signatures}

\subsection{Magnetism}
\label{magnetism}

There are several theories on the origin of magnetic fields in white dwarfs \citep{ferrario15,ferrario20}. One theory is that the field could be a fossil field remaining from the progenitor star if it was a highly magnetic Ap/Bp main sequence star \citep{tout04}, but the masses of these stars are typically too low to create the massive white dwarfs in our hot/massive subsample. Therefore, remnant fossil fields are unlikely the origin of the magnetic fields in our sample. 

Another possibility is a crystallization induced dynamo generating a magnetic field \citep{isern17}. 
\citet{bagnulo22} suggest that magnetic fields that appear in older, lower mass white dwarfs may likely be due to a crystallization induced dynamo.
However, the delay between crystallization and the magnetic field break out is on the order of a few billion years \citep{blatman24}. 
\citet{camisassa24} present an alternative hypothesis where the main-sequence dynamo fields emerge on the surface during the white dwarf phase.
\citet{bagnulo22} and \citet{moss25} find evidence of two separate populations of magnetic white dwarfs. The first is a population of young
and massive magnetic white dwarfs with field strengths indicative of a merger origin, such as expected from double white dwarf mergers
\citep{garcia12,briggs14}, and the second is older and average-mass white dwarfs with weaker fields that likely originate
from either a crystallization-induced dynamo or a main-sequence dynamo that surfaces during the white dwarf stage. 

Our sample is made up of hot ($T_{\rm eff}\geq11,000$ K) and massive ($M\geq0.9~M_\odot$) white dwarfs. The average age of this hot/massive sample is 0.80 Gyr. 
We find 67 magnetic white dwarfs in the hot/massive sample and 2 hot DCs that have to be magnetic. 
Several of these magnetic white dwarfs show merger signatures of rapid rotation or high tangential velocity.
Hence, a merger event is the most likely origin for these strongly magnetic, young, and hot white dwarfs \citet{garcia12,briggs14,bagnulo22,moss25}. 

\begin{figure}
\centering
\includegraphics[width=3.4in, clip, trim={0in 0in 0.4in 0.4in}]{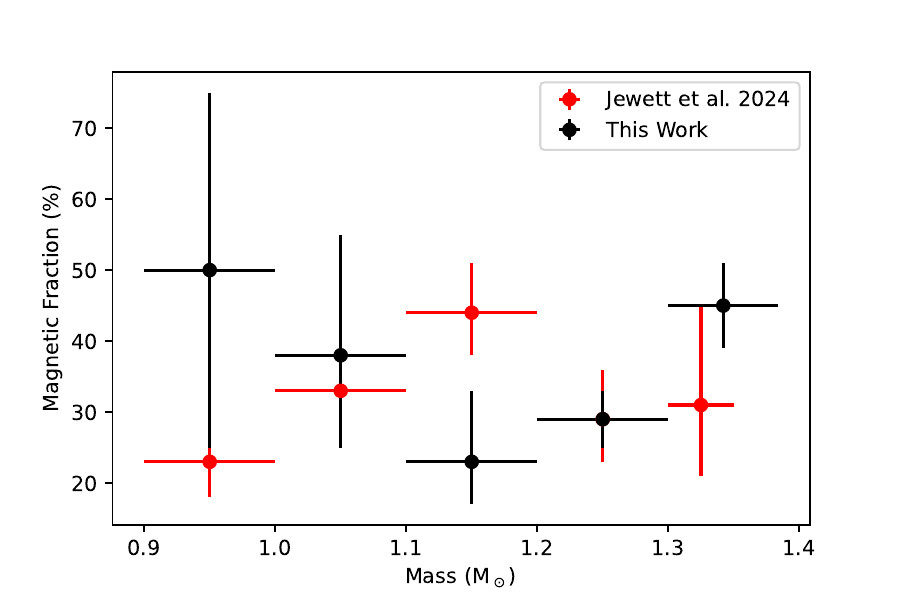}
\caption{A comparison of the magnetic fractions of the 100 pc massive white dwarf sample of \citet[][red points]{jewett2024} and this study (black points). The latter provides superior constraints above $1.2~M_\odot$ due to the significantly larger sample size.}
\label{fig:magnetic fraction}
\end{figure}

\begin{figure*}
\hspace{-0.4in}
\includegraphics[width=2.6in, clip, trim={0in 0in 0in 0in}]{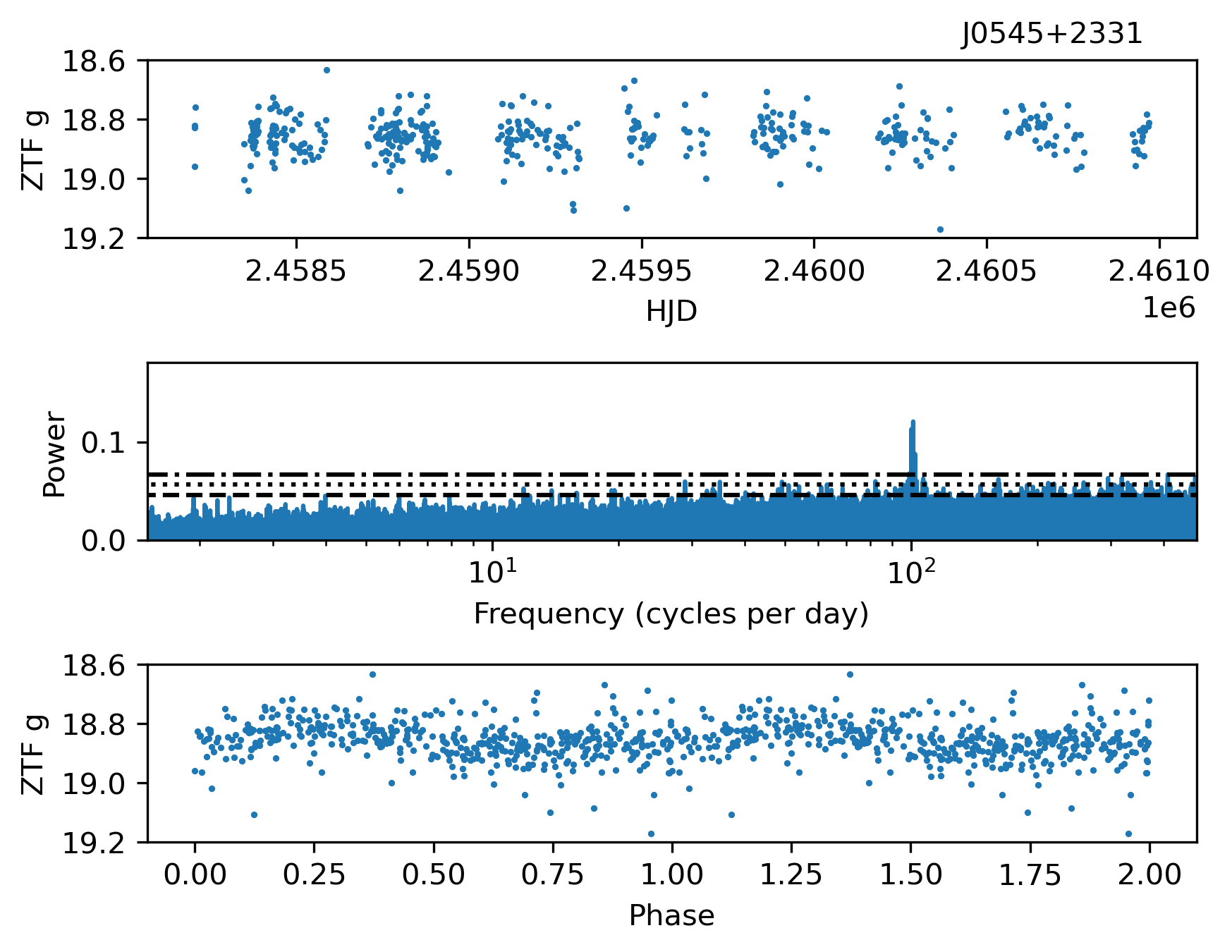}
\includegraphics[width=2.6in, clip, trim={0in 0in 0in 0in}]{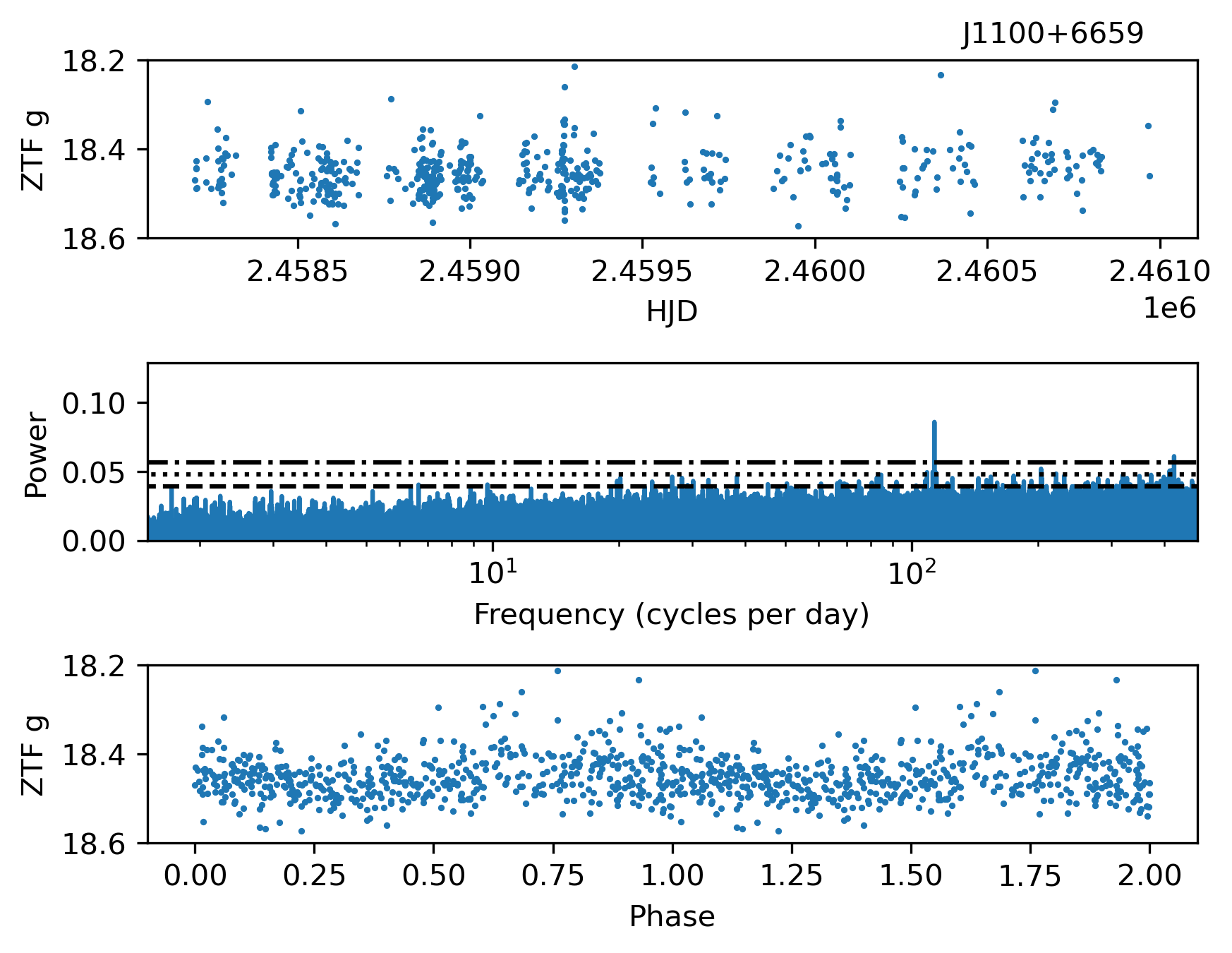}
\includegraphics[width=2.6in, clip, trim={0in 0in 0in 0in}]{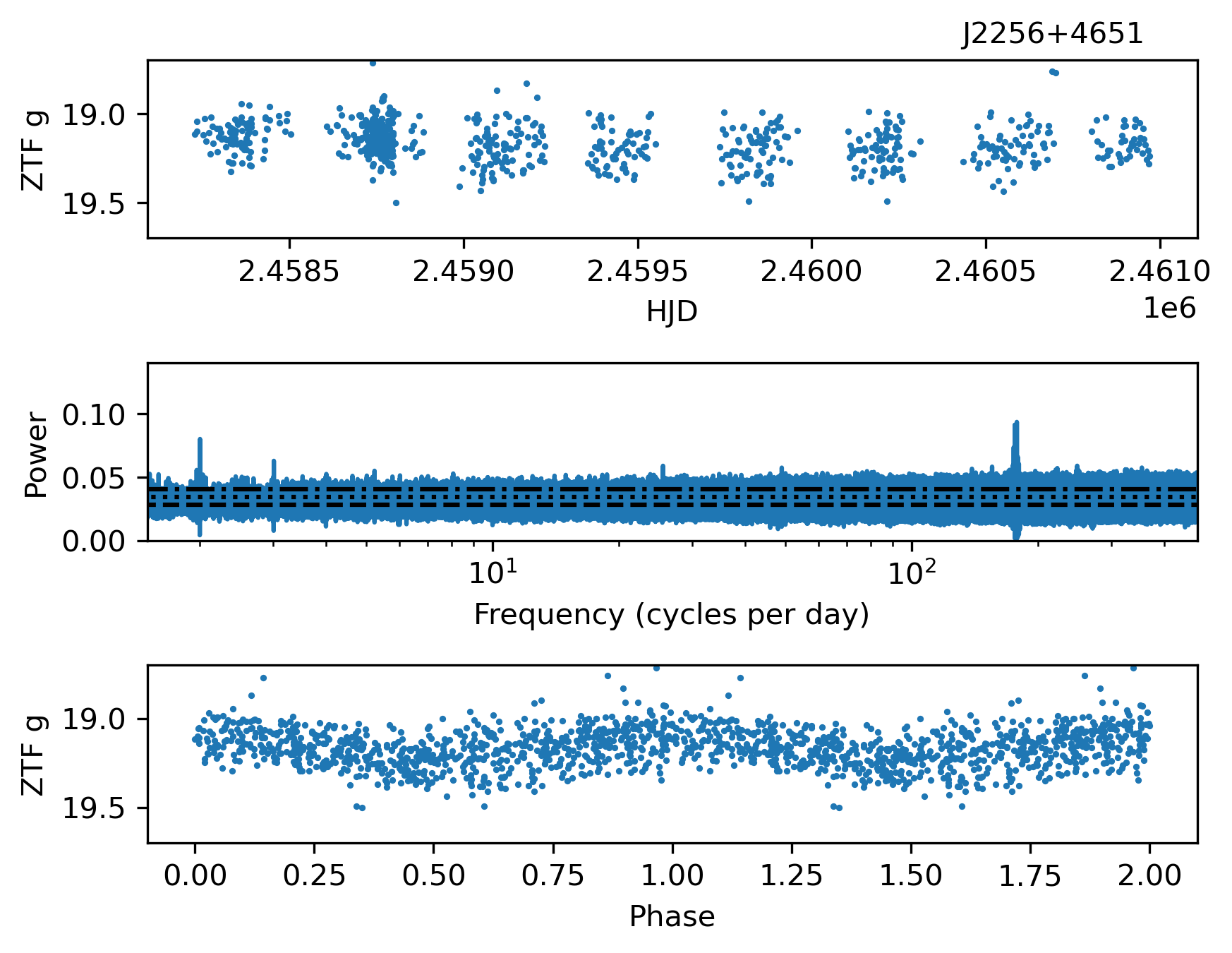}
\caption{The ZTF light curves (top) and Lomb-Scargle periodograms (middle panels) for J0545+2331, J1100+6659, and J2256+4651. The dashed lines in the middle panels show the 10\%, 1\%, and 0.1\% false alarm levels (from bottom to top). The bottom panels show the phased light curves using the highest peaks
detected in the periodograms.}
\label{fig:ztf}
\end{figure*}

Figure \ref{fig:magnetic fraction} shows the magnetic fraction as a function of mass for our sample (black points). We also show the magnetic fractions
obtained from the 100 pc massive white dwarf sample of \citet[][red points]{jewett2024} for comparison. The fraction of magnetic stars ranges from about
23\% to 50\% over the mass range shown here. Because our sample is biased towards more massive objects with $M\geq1.2~M_\odot$, the number of stars at lower masses is relatively small. Therefore, the error bars in the magnetic fraction calculations are relatively large for stars with $M<1.2~M_\odot$.
Given the errors, we find an overall agreement between the results from the 100 pc sample and this study. For the 1.1-1.2 $M_\odot$ bin, the magnetic fraction is much lower compared to the 100 pc sample. However, given the relatively large measurement errors, this difference is only $1.7\sigma$ significant, and it is likely due to the different selection biases present in our survey (which is biased towards higher masses) and the 100 pc sample.  

The most significant improvements in the magnetic fraction are for the 1.2-1.3 and $M\geq1.3~M_\odot$ bins. The 100 pc massive white dwarf sample of \citet{jewett2024} comprises of 49 and 13 stars in those mass bins, respectively, whereas our sample includes 112 and 60 stars in the same mass ranges.
Hence, the results from this survey are much more constraining than before; we find magnetic fractions of $29_{-4}^{+4}$\% and $45_{-6}^{+6}$\% for
1.2-1.3 and $M\geq1.3~M_\odot$, respectively. The latter is consistent with, but much more precise than $31_{-10}^{+14}$\% as derived from the 100 pc sample.

\subsection{Rotation}

Average mass white dwarfs (likely formed via single stellar evolution), $0.51 \le M\le0.73~M_\odot$, have on average 35 hour rotation periods \citep{hermes17}. Massive white dwarfs that form through single stellar evolution may have shorter rotation periods. For example, the pulsating ultramassive white dwarf WDJ004917.14$-$252556.81 does not show any merger signatures, and has a tentative rotation period of 7-16 h \citep{caliskan25}.
On the other hand, massive, magnetic white dwarfs typically have rotation periods shorter than $\sim$10 hours \citep{kawka20}. \citet{schwab21} predicts
merger products to have rotation periods between $\sim$10 and 20 minutes due to the angular momentum from the initial white dwarf binary system becoming part of the merger product's angular momentum.

Previously reported rotation periods from \citet{jewett2024}, include the hot/massive and likely magnetic DC white dwarfs J0327+2227, J0707+5611, and the magnetic J0718+3731 with periods 58.6, 63, and 11.27 min. These are all very fast rotation periods for white dwarfs. J0118$-$0156 (WDJ011810.31$-$015612.23), is another magnetic DAH
also reported to be rotating rapidly with a period of 54.69 min. These white dwarfs are likely merger remnants. 

For the rest of the hot/massive subsample, we search in the Zwicky Transient Facility \citep[ZTF,][]{bellm19} data for variability due to rapid rotation. We identify rapid rotation in three new white dwarfs; Figure \ref{fig:ztf} shows the ZTF $g$ band light curves (top panels), the Lomb-Scargle diagrams
(middle), and phase folded light curves (bottom panels) for these stars.
J0545+2331 (WDJ054542.85+233123.10) is a magnetic DA white dwarf with $M=1.251_{-0.022}^{+0.021}~M_\odot$ and $T_{\rm eff}=22,050\pm1438$ K. Based on the ZTF light curve,
we detect significant variability at a rotation period of 14.4 minutes. Because this is a magnetic white dwarf, and it is rapidly rotating, it is likely
a merger product. 

The middle panels in Figure \ref{fig:ztf} show the ZTF data for the DA white dwarf J1100+6659 (WDJ110036.98+665949.11). This star has $M=1.305_{-0.006}^{+0.005}~M_\odot$ and
$T_{\rm eff}=19,427\pm488$ K. The ZTF data reveals a fast 12.96 minute period. J1100+6659 is too hot to be in the ZZ Ceti instability strip. Hence,
the variability must be due to rapid rotation, which again indicates a merger remnant. 
Finally, the right panels in Figure \ref{fig:ztf} show the ZTF data on J2256+4651 (WDJ225607.09+465102.05), which is also a magnetic DAH white dwarf with
$M=1.009_{-0.071}^{+0.072}$ and $T_{\rm eff}=18,957\pm2094$ K. This white dwarf has a rotation period of 8.64 minutes. This is another massive, magnetic,
and hot white dwarf that shows rapid rotation, making it likely the product of a merger.

\subsection{Kinematics}

We calculate the tangential velocity for each white dwarf using the Gaia DR3 proper motions and \citet{bailerjones21} distances. Tangential velocity
is a useful discriminator for older stars, i.e. thick disk and halo members, that are kinematically heated. Ultramassive white dwarfs that were formed in the
thick-disk or halo would have cooled below 11,000 K by now, and they should not be present in our sample if they formed through single stellar evolution and if they had sufficiently large progenitor masses. However, such white dwarfs could appear young and hot, if they had their cooling age reset by a merger event, or if they suffer from long cooling delays due to $^{22}$Ne distillation \citep{bedard24, cheng19}.

\begin{figure}
\centering
\includegraphics[width=3.4in, clip, trim={0in 0in 0.4in 0.4in}]{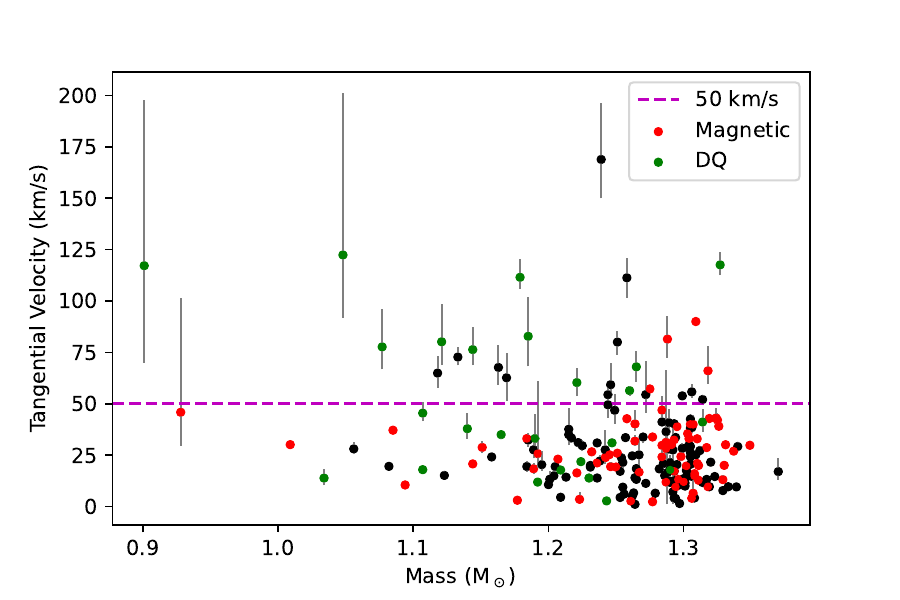}
\caption{Tangential velocity versus mass for the hot/massive white dwarfs in our sample. Green points mark DQs, red points mark magnetic white dwarfs, and black points mark the rest of the massive, hot white dwarfs. The dashed magenta line shows the 50 km s$^{-1}$ limit from \cite{wegg12}.}
\label{fig:vtan}
\end{figure}

Figure \ref{fig:vtan} shows the tangential velocities with the magnetic white dwarfs in red, the DQ white dwarfs in green, and the rest of the sample in black. Excluding the 3$\sigma$ outliers, the average tangential velocity of the sample is $\sim$25 km s$^{-1}$, which is consistent with the thin disk population. \citet{wegg12} used tangential velocities higher than 50 km s$^{-1}$ as the ``smoking gun'' signature of being a merger remnant. 
There are 28 white dwarfs in the hot/massive subsample that have tangential velocities above this limit. Out of the 28 high tangential velocity stars,
11 are DQs and four are magnetic. The highest tangential velocity of 168 km s$^{-1}$ is observed for a DA white dwarf with cooling age of 0.8 Gyr. This object is likely one whose cooling age was reset due to a merger event. 

The tangential velocities for the warm DQ population reveal that not all merger products have a high tangential velocity. The presence of DA white dwarfs above the 50 km s$^{-1}$ limit demonstrate that merger products do not have to have unusual atmospheres; they can also appear as DA white dwarfs.    
Recently, \citet{zamora26} reported that 39.4 \% of the warm DQs in their sample have velocities above 50 km s$^{-1}$. \citet{kilic25} found 54\% of their warm DQs were also above this velocity limit, and \citet{coutu19} found a similar percentage of 45\%. Our sample has 11 out of the 25 warm DQs, or 44\%, with tangential velocity above 50 km s$^{-1}$. This fraction is in agreement with what past surveys have found for the warm DQ population, providing further evidence that not all merger remnants have high tangential velocity.

\section{Discussion}
\label{discussion}

\subsection{The Merger Fraction}
\label{merger fraction}

\begin{deluxetable}{cccc>{\color{orange}}c>{\color{orange}}c>{\color{orange}}c}
\tabletypesize{\scriptsize}
\tablecolumns{7} \tablewidth{0pt}
\tablecaption{Number of magnetic white dwarfs and likely merger products as a function of mass (also shown are the calculated fractions).
We also include the numbers from the 100 pc sample \citep{jewett2024} for comparison. 
Our current sample provides superior constraints for the two highest mass bins.}
\label{tab:mergers} 
\tablehead{
\colhead{Mass} & \colhead{Total} & \colhead{Magnetic} & \colhead{Merger} &  \colhead{Total} & \colhead{Magnetic} & \colhead{Merger} \\
($M_{\odot}$) & stars & stars & products &  \colhead{stars} & \colhead{stars} & \colhead{products} \\
 &  & (This work) & &   & \colhead{(\citealt{jewett2024})} &
}
\startdata
$0.9-1.0$ & 2 &  1 ($50_{-25}^{+25}$\%) & 2 ($100_{-46}^{+}$\%) &  48 & 11 ($23_{-5}^{+7}$\%) & 12 ($25_{-5}^{+7}$\%) \\
$1.0-1.1$ & 8 &  3 ($38_{-13}^{+17}$\%) & 6 ($75_{-19}^{+9}$\%) &  36 & 12 ($33_{-6}^{+9}$\%) & 13 ($36_{-7}^{+9}$\%)\\
$1.1-1.2$ & 26 & 6 ($23_{-6}^{+10}$\%) & 21 ($81_{-10}^{+5}$\%) &  58 & 25 ($43_{-6}^{+7}$\%) & 45 ($78_{-7}^{+4}$\%) \\
\hline
$1.2-1.3$ & 112 & 32 ($29_{-4}^{+4}$\%) & 48 ($43_{-5}^{+5}$\%) &  49 & 14 ($29_{-6}^{+7}$\%) & 24 ($49_{-7}^{+7}$\%) \\
$1.3+$  & 60 & 27 ($45_{-6}^{+6}$\%) & 32 ($53_{-6}^{+7}$\%) &  13 &  4 ($31_{-10}^{+14}$\%) & 5 ($38_{-11}^{+15}$\%) \\
\hline
Sample    & 208 & 69 ($33_{-3}^{+4}$\%) & 109 ($52_{-3}^{+4}$\%) &  204 & 66 ($32_{-3}^{+4}$\%) & 99 ($49_{-4}^{+3}$\%)\\
\enddata
\end{deluxetable}

Our goal in this section is to determine what fraction of ultramassive white dwarfs are likely merger products. We consider the hot/massive subsample
with $T_{\rm eff}\geq11,000$ K and $M\geq0.9~M_\odot$ for these calculations. Previous works relied on small sample sizes of white dwarfs with $M > 1.3~M_\odot$ \citep{kilic23merger,jewett2024}. Our survey has significantly increased this population to 60 white dwarfs, allowing us to get tighter constraints on the merger fraction for this population. To calculate the merger fraction, we divide our sample into $0.1~M_\odot$ mass bins. Table \ref{tab:mergers} lists the total number of stars, the magnetic fraction, and merger fraction for each mass bin. We also include the values from the 100 pc sample of \citet{jewett2024} for comparison. We use the binomial probability distribution to calculate the errors in the merger fraction for each mass bin \citep[see][]{burgasser03}.  

Determining which white dwarfs are likely merger products requires searching for merger signatures. As discussed in previous sections, we did a detailed search for stars with unusual atmospheric composition, strong magnetic fields, rapid rotation, and unusual kinematics. Rapid rotation and strong magnetic fields are generally attributed to double white dwarf mergers \citep{garcia12, schwab21}. While kinematics and unusual atmospheric composition can also indicate a merger origin, they do not require double white dwarf mergers. For example, \citet{shen23} predict the overabundance of delayed Q branch white dwarfs to be due to white dwarf + subgiant mergers, as double white dwarf mergers do not produce enough neutron rich impurities for $^{22}$Ne distillation to occur and cause dramatic cooling delays. 
Additionally, \citet{ouldrouis26} find their delayed Q branch white dwarfs lack strong magnetism and rapid rotation, leading to the conclusion that they are not likely the product of double white dwarf mergers. The merger fraction we calculate here is representative of all possible formation channels, including mergers on the main-sequence, post-main-sequence, and white dwarf phase. Nevertheless, \citet{temmink20} predict double white dwarf systems to be the main contributor of merger products in this high mass range. 

We include white dwarfs that show at least one merger signature as a likely merger product. However, there are several caveats in our merger fraction estimates. The first is the availability of rotation and kinematics information. We do not have rotation periods or radial velocities for the entire sample. Hence, additional merger remnants could be missed, especially for white dwarfs that show no other merger signatures. Our assumption of all magnetic white dwarfs being merger remnants could be false. Even though the young, high-mass, and high-field objects are usually attributed to double white dwarf mergers \citep{bagnulo22,moss25}, a core-convective dynamo that forms on the main-sequence and emerges on the surface during the white dwarf phase may also be responsible for a significant fraction of the magnetic white dwarfs \citep{camisassa24,moss26}. Hence, the merger fractions may be over-estimated. The full results of our search for merger signatures can be found in Table \ref{short phys}. Many of the targets in our sample show only one signature of being a merger remnant, though
there are several targets that show multiple signatures. For example, J0157$-$0515 (WDJ015756.65$-$051541.73) is a DQA that has a tangential velocity of
$\sim76$ km s$^{-1}$, J0805$-$1702 (WDJ080502.93$-$170216.57) is a DAH with a tangential velocity of $\sim90$ km s$^{-1}$, and J2256+4651 (WDJ225607.09+465102.05) is magnetic and rapidly rotating with a period of $\sim$8 min.

\begin{figure}
\centering
\includegraphics[width=3.4in, clip, trim={0in 0in 0.4in 0.4in}]{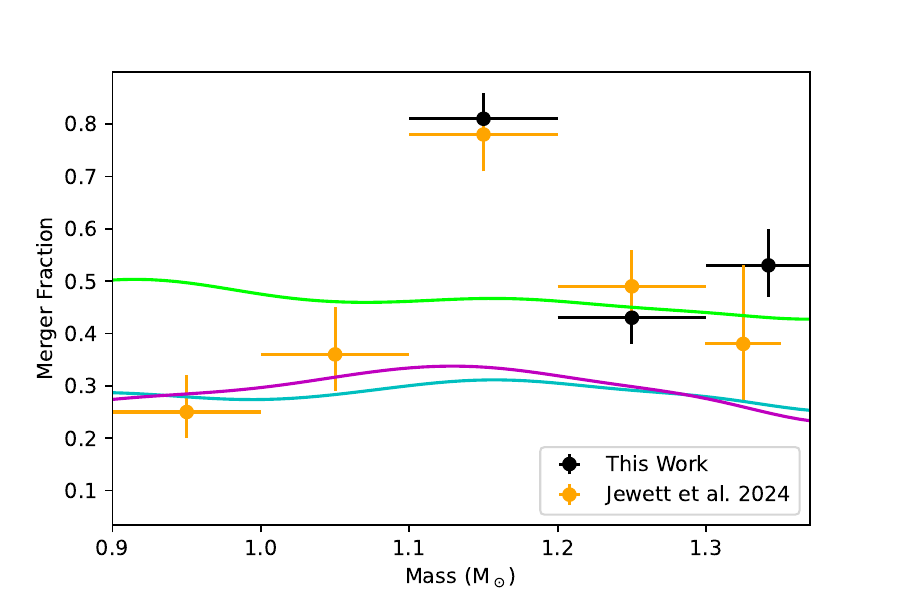}
\caption{The white dwarf merger fraction as a function of mass from this study (black) and the 100 pc sample \citet{jewett2024} (orange).  
The predicted merger fractions from \cite{temmink20} are shown for the DM91 (purple), $\alpha$-efficient (blue), and the $\alpha$-inefficient (green) models.}
\label{fig:mergerfraction}
\end{figure}

Figure \ref{fig:mergerfraction} shows the merger fractions as a function of mass from this study (black) and the 100 pc sample of \citet[][orange]{jewett2024}. Our sample includes only a few stars below $1.1~M_\odot$ (see Table \ref{tab:mergers}), hence the merger fractions for the 0.9-1.0 and
1.0-1.1 $M_\odot$ mass bins have relatively large errors that are not statistically meaningful. We exclude those measurements from this figure. On the other hand, for the two highest
mass bins the sample sizes are $\sim2$-4 times larger than \citet{jewett2024}. This is due to the fact that the search criteria we used targeted the most massive white dwarfs in Gaia.
Hence, our sample provides superior constraints for the most massive bins, including 112 stars with 1.2-1.3 $M_\odot$ and 60 stars more massive than $1.3~M_\odot$. These should be compared to 49 and 13 stars in the same mass bins in the 100 pc sample of \citet{jewett2024}. There is a remarkable
agreement between the merger fractions obtained from this work and the 100 pc sample, giving us confidence that these results are consistent. The merger
fractions are close for the 1.1-1.2 $M_\odot$ mass bin in both studies which is the mass range that is approximately twice the canonical white dwarf mass. Our estimates of $43_{-5}^{+5}$\% and $53_{-6}^{+7}$\%
merger fractions for the 1.2-1.3 $M_\odot$ and $M>1.3~M_\odot$, respectively, are more precise. The latter should be compared to $38_{-11}^{+15}$\%
reported in \citet{jewett2024}.

Figure \ref{fig:mergerfraction} also includes the predictions from the binary population synthesis models of \citet{temmink20} which takes into account multiple formation channels (see their Figure 10 for the contribution of each channel to the merger fraction). Here we include
the predictions from three different models, including the DM91, $\alpha$-efficient, and the $\alpha$-inefficient models
\citep[see][for details]{temmink20}. The difference in the population synthesis models are due to the input physics and the initial conditions used.
The DM91 model uses periods from a log-normal distribution and is otherwise similar to their default model \citep{duquennoy91}. The $\alpha$-efficient and $\alpha$-inefficient models make two different assumptions about the common envelope evolution using the alpha prescription. The $\alpha$ prescription is the canonical prescription where $\alpha$ represents the fraction of orbital energy used to unbind the envelope, where a larger $\alpha$ is more efficient \citep{temmink20}. Efficient $\alpha$ means more orbital energy can go into unbinding the common envelope before they merge so more binary systems can survive this phase of their evolution. Inefficient $\alpha$ has the opposite effect where more mergers will occur, hence the larger fractions predicted for this model. 
 
One of the most significant findings in this figure is the remarkably high merger fraction of 1.1-1.2 $M_\odot$ white dwarfs. We calculate a fraction of
$81_{-10}^{+5}$\%, whereas \citet{jewett2024} found a merger fraction of $78_{-7}^{+4}$\% for the same mass bin. Hence, this trend seems consistent; the merger
fraction for the 1.1-1.2 $M_\odot$ white dwarfs is much higher than predicted from the population synthesis models. 

\begin{figure}
\centering
\includegraphics[width=3.4in, clip, trim={0in 0in 0.4in 0.4in}]{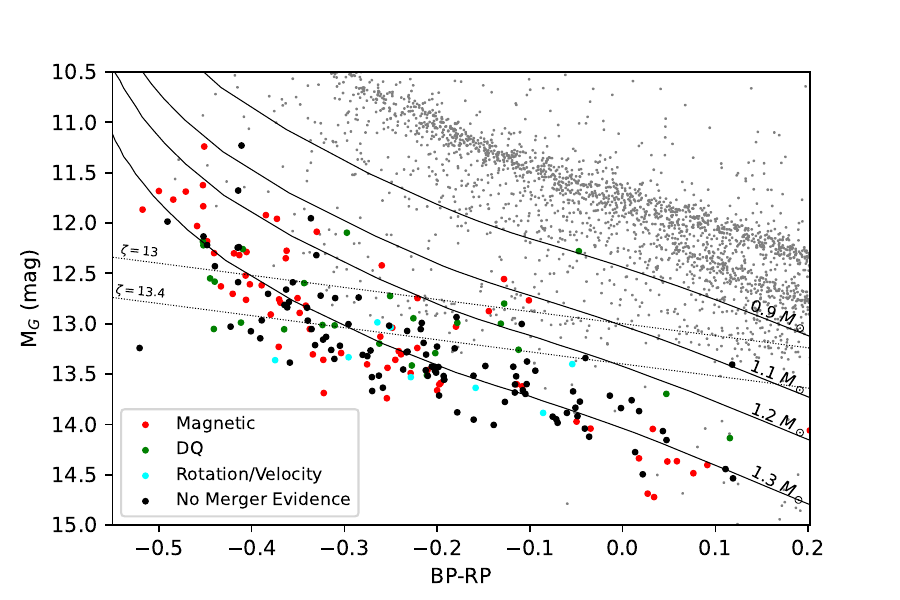}
\caption{Color magnitude diagram of the hot ($T_{\rm eff}>11,000$ K) and massive ($M>0.9~M_\odot$) white dwarfs in our sample. The solid lines show
the evolutionary tracks for 0.9, 1.1, 1.2, and $1.3~M_{\odot}$ pure H atmosphere white dwarfs \citep{bedard20}. Red, green, and cyan points
mark the magnetic, warm DQs, and white dwarfs with high tangential velocities or rapid rotation, respectively. The black points mark the rest of the hot/massive subsample that do not show any merger evidence.
The gray points in the background mark the 100 pc sample of white dwarfs. The Q branch is marked using lines of constant $\zeta$ \citep{camisassa21}.}
\label{fig:zetahr}
\end{figure}

Figure \ref{fig:zetahr} shows our hot and massive white dwarf sample in a color magnitude diagram with the Q-branch defined between
$\zeta=13$  and 13.4, where $\zeta = M_G - 1.2 \times (BP-RP)$ \citep{camisassa21}. Magnetic stars are marked in red, warm DQs in green, white dwarfs with high tangential velocity or rapid rotation in cyan, and the rest of the sample in black that show no merger evidence. Nineteen of the 25 DQs have masses in the 1.1-1.3 $M_\odot$ range. In addition, more than 1/3 of the white dwarfs in the 1.1-1.2 $M_\odot$ bin are DQs, including 10 DQs on the Q branch. The 1.1-1.3 $M_\odot$ range contains the most Q branch stars in our sample due to the temperature criteria that defines our sample. All warm DQs are found on or close to the the crystallization sequence \citep{kilic25}, most likely because they are stuck there due to significant cooling delays from distillation of neutron-rich impurities like $^{22}$Ne \citep{blouin21}. Therefore, the
inclusion of large numbers of DQ white dwarfs in our sample inflates the merger fraction. Based on the calculations by \citet{bedard24}, $^{22}$Ne
distillation approximately doubles the merger fraction for ultramassive white dwarfs. Hence, the inconsistencies between the very high merger fraction
observed for 1.1-1.2 $M_\odot$ white dwarfs and the population synthesis calculations are likely due to the effects of $^{22}$Ne distillation on
the local white dwarf population. 

Cooling delays from distillation also inflate the merger fractions for the two highest mass bins. Though we do not know the fraction of white dwarfs that undergo distillation for these mass bins, for comparison we can also estimate the merger fraction by only considering the stars below the Q branch. For the $1.2-1.3~M_\odot$, this reduces the sample size to 67 stars and the merger fraction to $33_{-5}^{+6}$\%, lower than the $43_{-5}^{+5}$\% estimated above. Considering the stars below the Q branch does lower the merger fraction making it clear that distillation plays a part in inflating the fraction for this mass range, though it does not appear to be inflated as much as the $1.1-1.2~M_\odot$ range. For the highest mass bin, we do not find a significant difference in the merger fraction if we only consider the stars below the Q branch.

Combining the results from the 100 pc sample for masses below $1.2~M_\odot$ and this work for masses above that limit, the merger fraction increases from
$\sim30$\% at $M\sim1~M_\odot$ to $\sim80$\% between 1.1-1.2 $M_\odot$, and then decreases to $\sim50$\% for the masses above $1.2~M_\odot$.
The overall merger fraction for our entire sample of $M>0.9~M_\odot$ white dwarfs is $52_{-3}^{+4}$\%. This is on the high end of what is predicted for $M>0.9 M_\odot$ white dwarfs by the binary population synthesis models \citep{temmink20}.
The observational data do not favor a single BPS model. At the lower mass end, the $\alpha$-efficient model is favored, whereas at the higher mass bins, a
merger fraction of $\sim$50\% favors the $\alpha$-inefficient model. However, given the inflated merger rates at the high mass end due to distillation,
it is difficult to pick a BPS model without additional constraints on the impact of distillation on the merger fraction of the highest mass white dwarfs.

\subsection{An Empirical Upper Mass Limit for White Dwarfs}
\label{mass limit} 

One of the best examples of an ultramassive white dwarf that likely formed through single star evolution is J0238+7642 (Gaia DR3 560883558756079616),
which is also included in our sample. This is an ultramassive white dwarf suspected of being an escapee from the Hyades Cluster with spectroscopic
parameters $T_{\rm eff} = 26,400 \pm 200$ K and $M = 1.317^{+0.014}_{-0.018}~M_{\odot}$ assuming an ONe core and a cooling age of $556^{+15}_{-22}$ Myr
\citep{miller23}. Our investigation of this white dwarf also shows no evidence of a merger origin. We obtain $T_{\rm eff} = 22,039_{-1115}^{+1114}$ K,
$M = 1.305_{-0.013}^{+0.012}~M_{\odot}$ (assuming a CO core), and cooling age $0.72_{-0.08}^{+0.09}$ Gyr from the photometric method. 

The most massive white dwarf in the Gaia sample with no obvious evidence of a merger origin is J1650$-$4752 (WDJ165019.13$-$475208.84) with $M = 1.340 \pm 0.039~M_{\odot}$ (assuming a CO core, and including GR effects) or $M=1.320 \pm 0.035~M_\odot$ (for an ONe core).
One can never rule out a binary origin, but this star appears to be a normal DA white dwarf with a tangential velocity of $~29$ km s$^{-1}$. 


\begin{figure}
\centering
\includegraphics[width=3in, clip=true, trim=0.4in 0.8in 0.2in 1.1in]{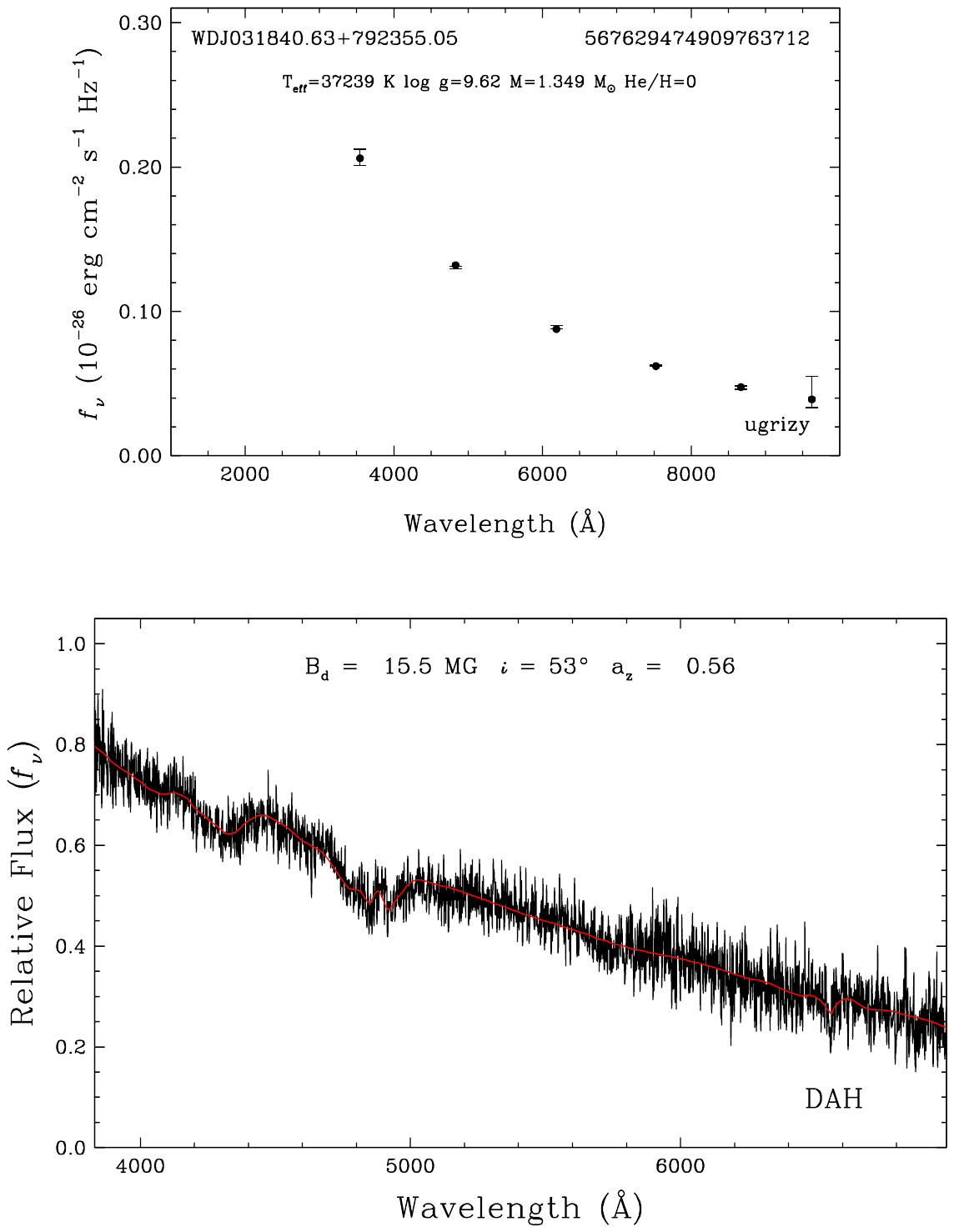}
\caption{Model fit to the most massive white dwarf in our sample. J0318+7923 is a magnetic white dwarf with $B=15.5$ MG. The best-fitting model
indicates $M = 1.349 \pm 0.006~M_\odot$ including GR effects.}
\label{figj0318}
\end{figure}

\begin{figure}
\centering
\includegraphics[width=3.4in, clip, trim={0in 0in 0.4in 0.4in}]{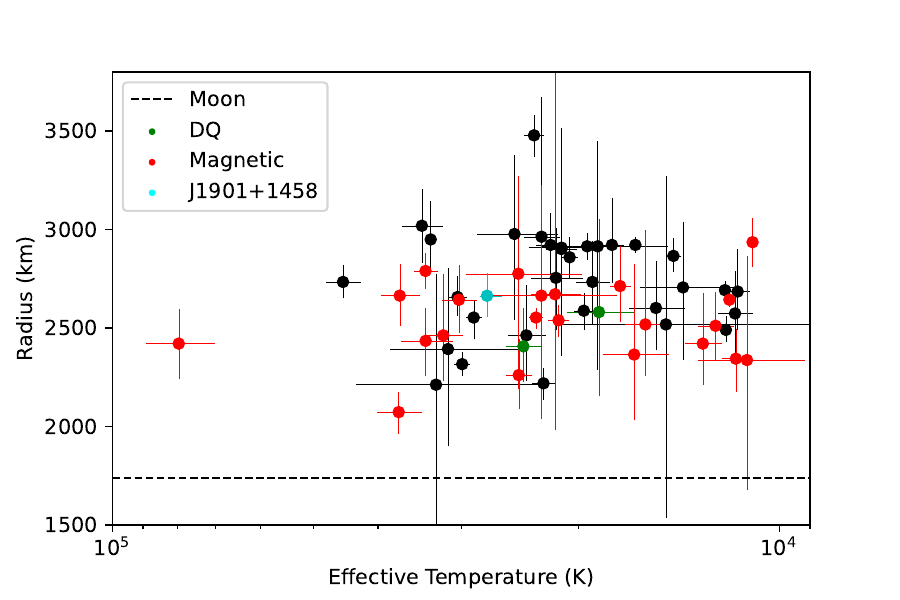}
\caption{Radius versus effective temperature for the hot and $M\geq1.3~M_\odot$ white dwarfs. The DQ white dwarfs are marked in green, and the magnetic white dwarfs are in red. The rest of the sample are marked in black. J1901+1458 is shown in blue using parameters from \citet{desai26}. The dashed black line represents the radius of the moon for comparison.}
\label{fig:radius}
\end{figure}

\begin{figure}
\centering 
\includegraphics[width=3.4in]{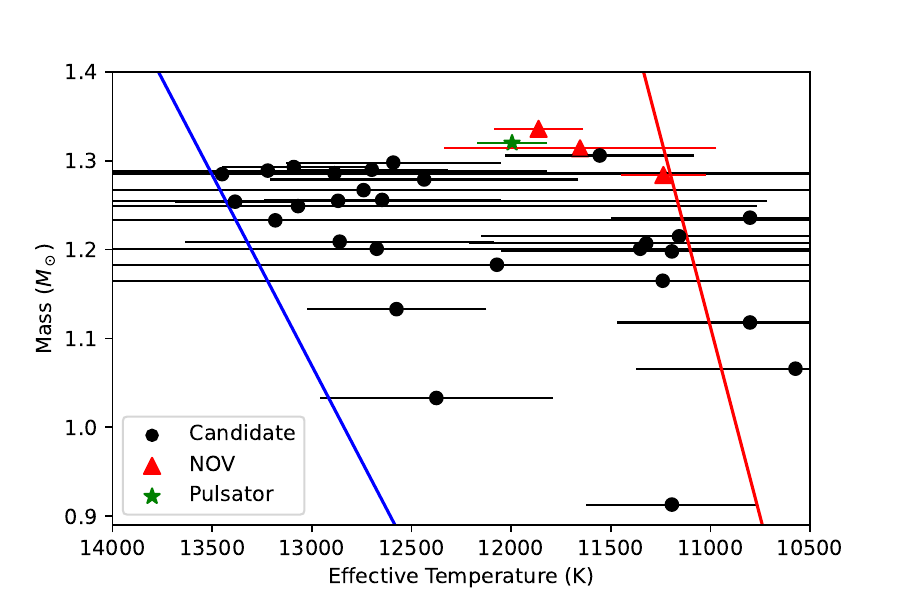}
\caption{Photometric masses and effective temperatures for massive DA white dwarfs in or near the ZZ Ceti instability strip. The candidates near the strip are restricted to stars with effective temperature errors less than 1000 K. The blue and red lines mark the empirical boundaries of the photometric instability strip from \citet{vincent20}. The black points represent the white dwarf pulsator candidates. The green star marks a known pulsator in the sample: J0959$-$1828, and the red triangles mark non-variable stars (NOVs): J0127$-$2436 and J1140+2322 from \citet{jewett2025} and J1604$-$2434 (this study).}
\label{fig:zzceti}
\end{figure}

Among the merger remnants, the previous record holder in mass was J1901+1458 (ZTF J190132.9+145808.7) with $M=1.32~(1.29)~M_\odot$ for a CO (ONe) core \citep{caiazzo21,desai26}. This white dwarf shows rapid rotation with a period of 6.94 min and a magnetic field strength ranging between 600-900 MG \citep{caiazzo21}.

Excluding J1627$-$7603, which has discrepant photometric and spectroscopic fits, the highest mass object in our sample (based on models including GR effects) is the magnetic white dwarf J0318+7923 (WDJ031840.63+792355.05; Gaia DR3 567629474909763712). This star was previously
classified as an uncertain DA white dwarf (DA: spectral type) based on a Gaia XP spectrum \citep{vincent24}. Figure \ref{figj0318} shows our MMT spectrum and the model fit for this star. J0318+7923 is a magnetic DAH with $T_{\rm eff} = 37,239 \pm 2915$ K, $M = 1.349 \pm 0.006~M_\odot$, and $B=15.5$ MG. For comparison, the mass assuming an ONe core is $M=1.329 \pm0.006~M_\odot$.
Based on this mass estimate (with the caveats mentioned above) and the strong magnetism, it is likely a merger remnant, making this white dwarf the most massive merger remnant candidate known. 


Another notably massive white dwarf is J1153+0056 (WDJ115305.54+005646.15). This star has been previously classified as a DQ by \citet{liebert03} and \citet{dufour08} based on an SDSS spectrum, though \citet{dufour08} assumed $\log{g}=8$ due to the lack of a parallax measurement. Given the Gaia parallax, we revise its physical
parameters to $T_{\rm eff} = 24,217 \pm 1528$ K and $M = 1.327 _{-0.024}^{+0.012}~M_{\odot}$ including GR effects in the mass calculation \citep[see also a fit to its DESI spectrum in][]{kilic26}.
In addition to J1153+0056, the only other previously known warm DQ with $M>1.3~M_\odot$ was J2057$-$4434 (WDJ205700.44$-$342556.37) with
$M=1.310_{-0.015}^{+0.017}~M_\odot$ \citep{kilic25}. This is the hottest and most massive DAQ known. We add a new warm DQ to the $M>1.3~M_\odot$ sample with J0749+3540 (WDJ074947.00+354055.47), which has $T_{\rm eff}=18,646\pm2186$ and $M=1.314_{-0.037}^{+0.029}~M_\odot$. 

Unlike the mass estimates that rely on evolutionary models, we also have radii measurements that come directly from our photometric fits. These do not depend on the white dwarf
evolutionary models, and provide an independent check on the unusually high masses of these objects. 
Figure \ref{fig:radius} shows the radii versus effective temperature for the hot, $M>1.3~M_\odot$ white dwarfs in our sample, including magnetic white dwarfs (red), and DQs (green). J1901+1458 is included for reference \citep{caiazzo21,desai26}. The star with the smallest radius is J0318+7923, which is also the highest mass object in our sample (as discussed above). J0318+7923 has a radius of $2073_{-111}^{+104}$ km, about 20\% larger than the Moon. J0318+7923 represents the smallest white dwarf currently known with well defined parameters.

\subsection{Pulsations}

\begin{figure*}
\centering
\includegraphics[width=0.9\linewidth]{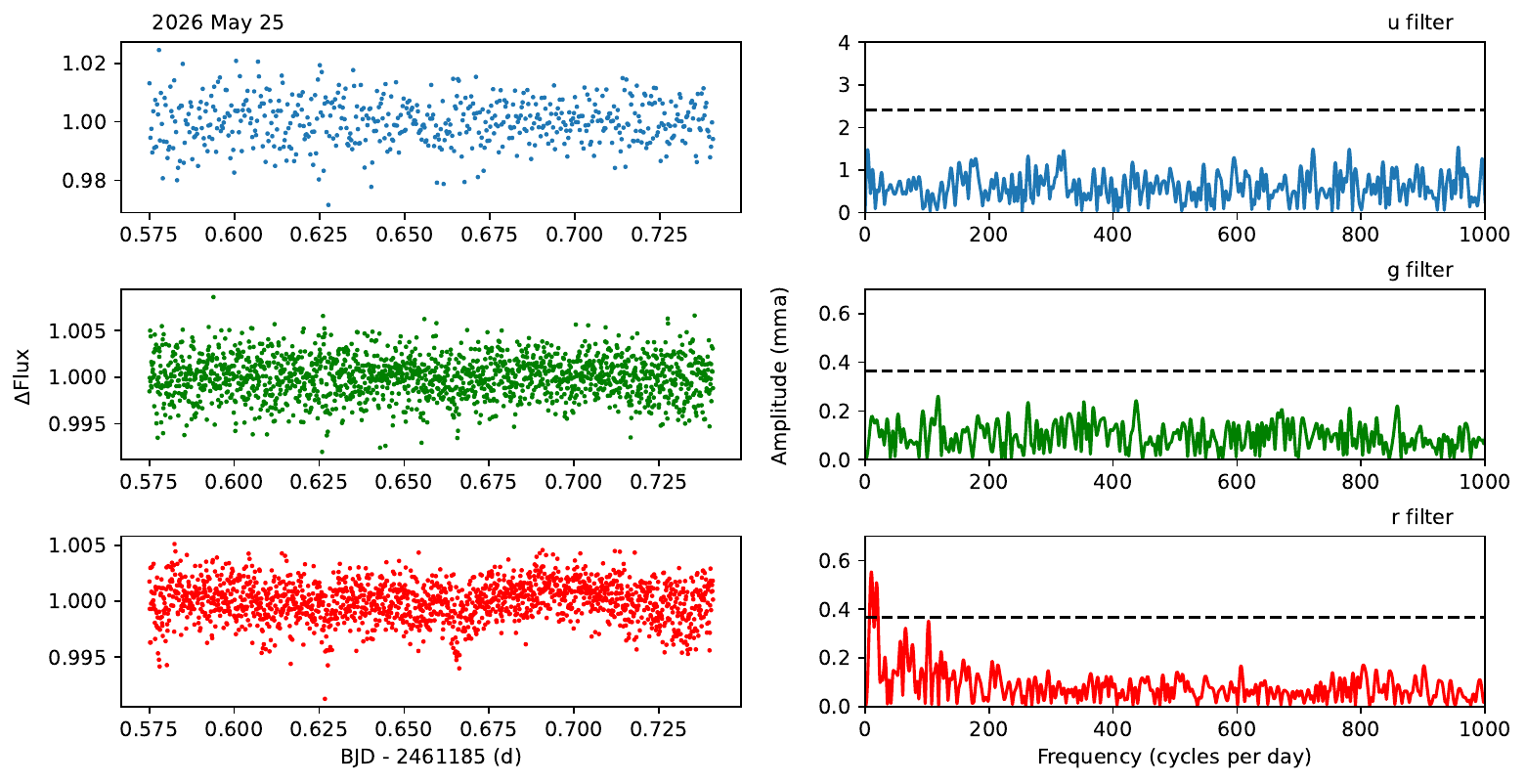}
\caption{Light curves and Fourier transforms for J1604$-$2434. Left panels show the light curves in the $u$, $g$, and $r$ filters. The right panels show the associated Fourier transform. The black dashed line in each panel shows the 4$\langle {\rm A}\rangle$ detection limit.}
\label{fig:j1604}
\end{figure*}

\begin{figure*}
\centering
\includegraphics[width=0.9\linewidth]{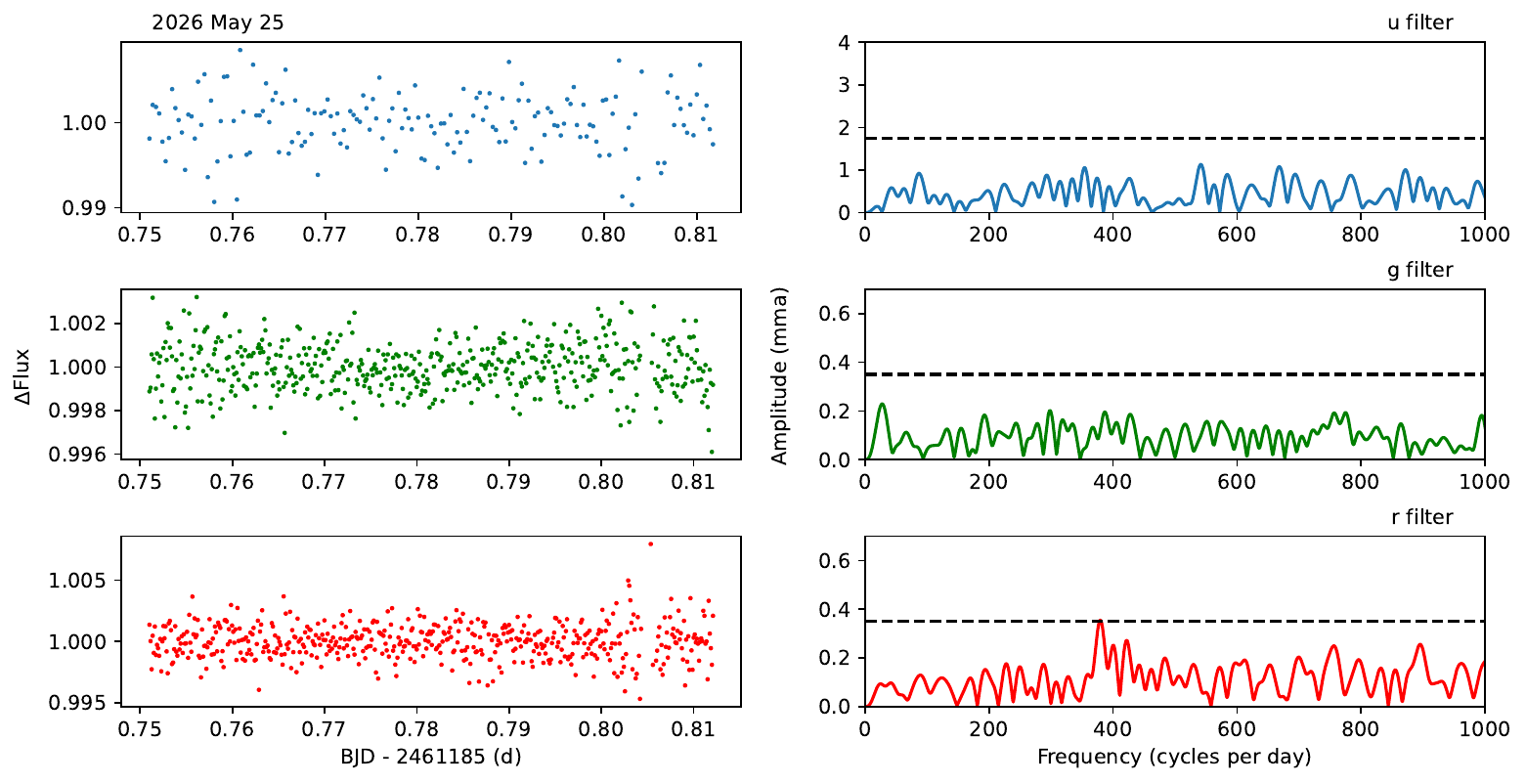}
\caption{Light curves and Fourier transforms for J1838$-$4120. The symbols are the same as in Figure \ref{fig:j1604}.}
\label{fig:j1838}
\end{figure*}

\begin{figure*}
\centering
\includegraphics[width=0.9\linewidth]{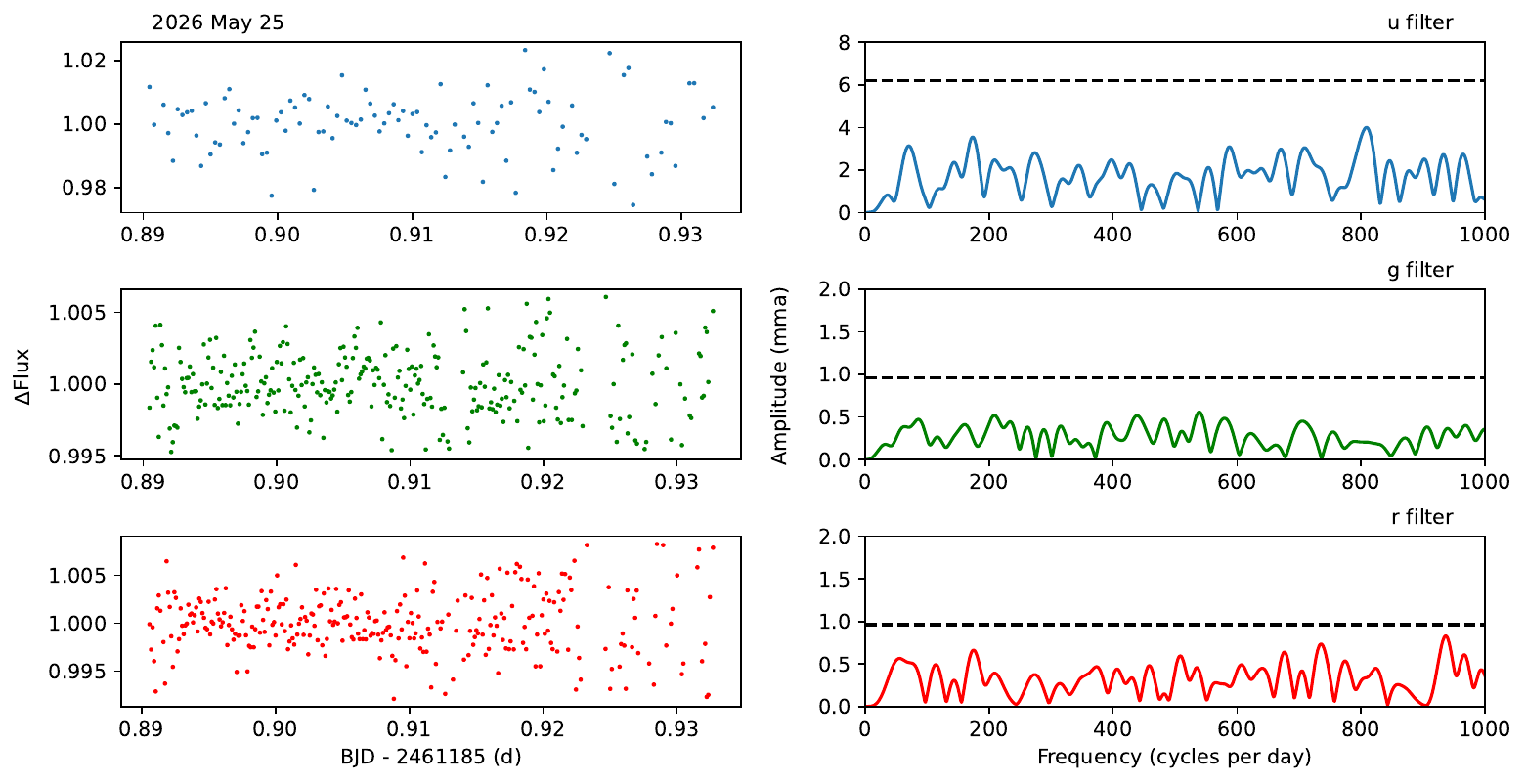}
\caption{Light curves and Fourier transforms for J2109$-$5206. The symbols are the same as in Figure \ref{fig:j1604}.}
\label{fig:j2109}
\end{figure*}

\begin{deluxetable*}{ccc}
\tabletypesize{\tiny}
\tablecolumns{3} 
\tablewidth{0pt}
\caption{The massive, hot DA pulsator candidates within or near the ZZ Ceti instability strip. This contains all white dwarfs within the photometric boundaries or near the boundaries with effective temperature errors less than 1000 K.}
\label{tab:zz_ceti_cand}
\tablehead{\colhead{Object Name} & \colhead{Phot. Mass} & \colhead{Phot. Temp}}
\startdata
J0145$-$1417 & $1.256_{-0.054}^{+0.038}$ & $12647_{-594}^{+593}$  \\
J0405+2119 & $1.133_{-0.042}^{+0.034}$ & $12575_{-451}^{+451}$  \\
J0452+3944 & $1.289_{-0.084}^{+0.039}$ & $13221_{-1403}^{+1403}$  \\
J0603$-$6526 & $1.255_{-0.086}^{+0.068}$ & $12868_{-2153}^{+2153}$  \\  
J0728$-$4633 & $1.165_{-0.244}^{+0.169}$ & $11239_{-2781}^{+2781}$  \\ 
J0729$-$2730 & $1.198_{-0.131}^{+0.066}$ & $11193_{-857}^{+857}$  \\ 
J0820$-$1608 & $1.201_{-0.156}^{+0.078}$ & $11352_{-873}^{+873}$  \\
J1106$-$5326 & $1.183_{-0.177}^{+0.126}$ & $12071_{-2207}^{+2208}$  \\ 
J1110+3047 & $1.293_{-0.061}^{+0.034}$ & $13090_{-362}^{+362}$  \\
J1229$-$5148 & $1.233_{-0.191}^{+0.101}$ & $13183_{-4912}^{+4912}$  \\ 
J1356+5423 & $1.279_{-0.051}^{+0.039}$ & $12436_{-774}^{+774}$  \\
J1447$-$6423 & $1.285_{-0.086}^{+0.049}$ & $13450_{-3991}^{+3992}$   \\  
J1452$-$2631 & $1.118_{-0.088}^{+0.075}$ & $10801_{-666}^{+666}$  \\
J1512$-$4731 & $1.201_{-0.191}^{+0.133}$ & $12674_{-4524}^{+4524}$ \\ 
J1518$-$5425 & $1.267_{-0.145}^{+0.067}$ & $12740_{-4215}^{+4215}$  \\ 
J1604$-$2434 & $1.314_{-0.017}^{+0.017}$ & $11654_{-684}^{+685}$  \\ 
J1612$-$6439 & $1.286_{-0.061}^{+0.048}$ & $12886_{-2636}^{+2637}$ \\ 
J1615+3019 & $1.306_{-0.018}^{+0.014}$ & $11555_{-473}^{+473}$  \\ 
J1643$-$4843 & $1.249_{-0.085}^{+0.071}$ & $13069_{-2303}^{+2302}$  \\
J1806$-$1703 & $1.298_{-0.016}^{+0.013}$ & $12591_{-540}^{+539}$  \\ 
J1911+1816 & $1.236_{-0.046}^{+0.042}$ & $10801_{-697}^{+697}$  \\ 
J1936$-$2513 & $1.209_{-0.046}^{+0.043}$ & $12860_{-773}^{+774}$ \\ 
J1942+2019 & $0.913_{-0.227}^{+0.170}$ & $11193_{-434}^{+433}$  \\ 
J2020+0504 & $1.033_{-0.186}^{+0.127}$ & $12375_{-583}^{+583}$  \\ 
J2042+3041 & $1.215_{-0.132}^{+0.089}$ & $11157_{-995}^{+995}$  \\ 
J2130$-$2526 & $1.066_{-0.455}^{+0.155}$ & $10573_{-799}^{+800}$  \\ 
J2238$-$0802 & $1.254_{-0.012}^{+0.014}$ & $13385_{-299}^{+299}$  \\ 
J2243+2108 & $1.207_{-0.117}^{+0.077}$ & $11322_{-890}^{+891}$ \\ 
J2353+3809 & $1.290_{-0.014}^{+0.012}$ & $12699_{-383}^{+382}$  \\ 
\enddata
\end{deluxetable*}

The temperature cut off we impose on our sample ($T_{\rm eff}\geq11,000$ K) overlaps with the photometric ZZ Ceti Instability strip. Hence, a byproduct
of our survey is to provide a list of ultramassive DA white dwarfs that may pulsate. Pulsating ultramassive white dwarfs provide a unique opportunity
to study their crystallized interiors and potentially constrain their core composition and the H envelope thickness. Asteroseismology of $M>1.3~M_\odot$
white dwarfs can also potentially constrain the impact of general relativity on the gravity (g)-mode period spectrum \citep{corsico23}. However, ultramassive pulsators are relatively rare \citep{kanaan92,hermes13}. Through follow-up time-series photometry of the massive white dwarfs in the
100 pc sample, \citet{jewett2025} identified 16 pulsators with $M>0.9~M_\odot$, including four objects more massive than $1.2~M_\odot$
\citep[see also][]{kilic23J0049,degeronimo25,caliskan25,uzundag25,ouldrouis26}. 

Table \ref{tab:zz_ceti_cand} presents a list of massive DA white dwarfs in our sample that are in or near the ZZ Ceti instability strip based on their photometric effective temperatures. Figure \ref{fig:zzceti} shows the photometric parameters of these DA white dwarfs along with the empirical boundaries of the instability strip. Due to the relatively uncertain distances, the uncertainties in the photometric parameters are relatively large. This leads to a large number of potential candidates near the ZZ Ceti strip, mainly hotter than the blue edge. We refine the sample of candidates by including those with effective temperatures within the boundaries of the photometric strip, and restrict the candidates near the strip to those stars with effective temperature errors less than 1000 K.

There are four white dwarfs within the photometric instability strip that have recent time-series photometry. One of these, J0959-1828 (WDJ095933.33$-$182824.16), is the most massive pulsating white dwarf currently known with mass $M= 1.32 \pm0.004~M_\odot$, assuming a CO core \citep{jewett2025}. The other three white dwarfs observed do not show evidence of variability. J0127$-$2436 (WDJ012734.59$-$243620.14) was previously discussed in \citet{jewett2025}, and was only observed once for a relatively short 1.7 hours. J1140+2322 (WDJ114012.83+232205.46) was previously discussed in \citet{kilic23merger} and was observed four times with no significant modes detected. 

We have started time-series photometric observations of the remaining candidates. We have so far observed J1604$-$2434 (WDJ160408.49$-$243405.17) with ULTRACAM \citep{dhillon07} on the 3.5m ESO New Technology Telescope (NTT). ULTRACAM observes simultaneously in the super-SDSS $u_s$, $g_s$, and $r_s$ filters and has negligible 24 ms dead-time between exposures \citep{brown22}. Figure \ref{fig:j1604} shows the light curves and corresponding Fourier transforms for J1604$-$2434 observed on UT 2026 May 25 using 8 second exposures. In order to set a detection limit for which peaks in the transform are significant, we use 4 times the average amplitude of the Fourier transform (4$\langle {\rm A}\rangle$) \citep{breger93,degeronimo25,caliskan25}. The Fourier transforms do not show any significant modes in the $u$ and $g$-bands, where pulsations should have the highest amplitudes. The $r$ band data does show two significant peaks at low frequencies at $\sim$10 and $\sim$18 cycles per day. These correspond to periods too long to be from pulsations, and therefore we classify this white dwarf as non-variable. There are several reasons why non-variable white dwarfs may be located in the strip. If the white dwarfs are weakly magnetic, then it's possible the field is suppressing pulsations. Another reason is that the pulsation modes could have very low amplitudes beyond the detection capabilities of the telescope used. And, the method used to determine the effective temperature can affect where the white dwarf is located in the strip. This a question of the spectroscopic versus photometric methods. For detailed discussion on these points, see \citet{jewett2025}. 

Figure \ref{fig:j1838} and Figure \ref{fig:j2109} show the ULTRACAM data on J1838$-$4120 (WDJ183805.36$-$412024.01) and J2109$-$5206 (WDJ210941.17$-$520628.10), respectively. These stars are hotter than the blue edge with large temperature errors potentially placing them within the photometric boundaries of the instability strip. J1838$-$4120 has relatively large temperature errors with $T_{\rm eff}=15,026 \pm5833$ K. J2109$-$5206 is a similar case with $T_{\rm eff}=14,284\pm4136$ K. We do not see pulsations in these white dwarfs because they lie outside
of the instability strip. These observations reaffirm our choice to limit our candidate sample to stars near the strip with effective temperature errors less than 1000 K.

\section{Conclusions}
\label{conclusions}

We present the results from a spectroscopic survey of hot and massive white dwarfs selected from
the Gaia DR3 white dwarf catalog of \citet{gentilefusillo21} with $T_{\rm eff}\geq11,000$ K and $M\geq1.3~M_\odot$. Our two main goals are 1) to provide a precise merger rate for the most massive white dwarfs, and 2) to find the most massive white dwarf in the solar
neighborhood. 
There are 293 candidates in \citet{gentilefusillo21} that satisfy our selection criteria based on pure H atmosphere model fits. Our follow-up
spectroscopy is 97\% complete. We use the observed spectra in combination with the Gaia distances and photometry from the SDSS, Pan-STARRS,
and Skymapper to refine the physical parameters of these objects. Our analysis reveals a sample of 208 hot/massive white dwarfs, with 25 warm DQs (including a new DAQ), two rare ultramassive DZs, and 66 magnetic white dwarfs.

Our sample includes 112 stars with $M=1.2-1.3~M_\odot$ and 60 stars with $M\geq1.3~M_\odot$. This sample is significantly larger than the massive white dwarf samples in the literature, and it provides superior constraints on the fraction of magnetic white dwarfs and merger remnants
for the most massive white dwarfs. We find a magnetic fraction of $29^{+4}_{-4}$\% and $45^{+6}_{-6}$\% for $M=1.2-1.3~M_\odot$ and $M\geq1.3~M_\odot$ white
dwarfs, respectively. The magnetic fractions determined for this work and for the 100 pc sample are independent from each other, yet consistent between the two samples. These new magnetic fractions are more precise than those obtained from the 100 pc sample \citep{jewett2024}.

We search for merger remnants in our sample based on unusual atmospheric composition (warm DQs with C atmospheres), magnetism, rapid rotation, and
high tangential velocity. We find an overall merger fraction of 52\% for $M>0.9~M_\odot$, which favors an inefficient common envelope based on the
binary population synthesis models from \citet{temmink20}. Interestingly, the merger fraction peaks at $81_{-10}^{+5}$\% for the 1.1-1.2 $M_\odot$ range, before coming down to $43_{-5}^{+5}$\% for 1.2-1.3 $M_\odot$ stars and $53_{-6}^{+7}$\% for $M>1.3~M_\odot$. \citet{jewett2024} also found a similarly high merger fraction for the 1.1-1.2 $M_\odot$ mass range, and attributed this to the over-representation of warm DQs that are suffering from long cooling delays due to $^{22}$Ne distillation. The two highest mass bins also likely have inflated merger fractions, and we set a lower limit $33_{-5}^{+6}$\% for the 1.2-1.3 $M_\odot$ range by only using stars below the Q branch.

The most massive white dwarf in our sample is the magnetic DAH J0318+7923, which has an estimated mass of $M=1.349\pm0.006~M_\odot$ under the
assumption of a CO core and $M=1.329\pm0.006~M_\odot$ assuming an ONe core including GR effects. Given the strong magnetism and the large mass, this is likely a merger remnant. Among the population of white dwarfs that
do not show any evidence of being a merger remnant (DA white dwarfs with no evidence of rapid rotation or high tangential velocity), J1650$-$4752
stands out with a mass of $1.340\pm0.039~M_\odot$. Hence, these stars provide an empirical measurement of the upper mass limit for stars that
can be supported through electron degeneracy pressure. However, there is a caveat in our mass measurements; the unknown core composition. If these white dwarfs have ONe cores, then their masses would be systematically lower (see Table \ref{short phys}). 

Further progress requires additional constraints on the core compositions of these stars. This can be achieved through asteroseismology as our sample criteria overlap with the ZZ Ceti instability strip. We find many new
ultramassive white dwarfs within or near the instability strip, and we are obtaining time-series photometry on them to search for pulsations. These
observations will expand the known ultramassive pulsator sample.

\begin{acknowledgments}
We thank the referee for their detailed report and valuable suggestions that have improved this paper. 

This work is supported in part by the NSF under grant  AST-2508429, the NASA under grants 80NSSC24K0380, and 80NSSC24K0436, the NSERC Canada, the Fund FRQ-NT (Qu\'ebec), and the Smithsonian Institution.

Based on observations obtained at the MMT Observatory, a joint facility of the Smithsonian Institution and the University of Arizona.
This paper includes data gathered with the 6.5 m Magellan Telescopes located at Las Campanas Observatory, Chile.

Based on observations collected at the European Organisation for Astronomical Research in the Southern Hemisphere under ESO program 117.29MD.

VSD and ULTRACAM are supported by STFC grant  ST/Z000033/1.

Based on observations obtained at the international Gemini Observatory, a program of NSF NOIRLab, which is managed by the Association of Universities for Research in Astronomy (AURA) under a cooperative agreement with the U.S. National Science Foundation on behalf of the Gemini Observatory partnership: the U.S. National Science Foundation (United States), National Research Council (Canada), Agencia Nacional de Investigaci\'{o}n y Desarrollo (Chile), Ministerio de Ciencia, Tecnolog\'{i}a e Innovaci\'{o}n (Argentina), Minist\'{e}rio da Ci\^{e}ncia, Tecnologia, Inova\c{c}\~{o}es e Comunica\c{c}\~{o}es (Brazil), and Korea Astronomy and Space Science Institute (Republic of Korea).

\end{acknowledgments}

\facilities{MMT (Blue Channel spectrograph), Gemini (GMOS), Magellan:Baade (MagE spectrograph), NTT - New Technology Telescope (ULTRACAM).}

\bibliography{references}{}
\bibliographystyle{aasjournalv7}

\end{document}